\documentclass[aps,prb,preprint,groupedaddress,showkeys]{revtex4-2}

\usepackage{graphicx}
\usepackage{subfigure}
\begin{document}


\title{Insights into electron transport in a ferroelectric tunnel junction}


\author{Titus Sandu}
\email[]{titus.sandu@imt.ro}
\affiliation{National Institute for Research and Development in Microtechnologies-IMT, 126A, Erou Iancu Nicolae Street, Bucharest, ROMANIA, 077190}
\author{Catalin Tibeica}
\affiliation{National Institute for Research and Development in Microtechnologies-IMT, 126A, Erou Iancu Nicolae Street, Bucharest, ROMANIA, 077190}
\author{Rodica Plugaru}
\affiliation{National Institute for Research and Development in Microtechnologies-IMT, 126A, Erou Iancu Nicolae Street, Bucharest, ROMANIA, 077190}
\author{Oana Nedelcu}
\affiliation{National Institute for Research and Development in Microtechnologies-IMT, 126A, Erou Iancu Nicolae Street, Bucharest, ROMANIA, 077190}
\author{Neculai Plugaru}
\affiliation{National Institute for Research and Development in Microtechnologies-IMT, 126A, Erou Iancu Nicolae Street, Bucharest, ROMANIA, 077190}


\date{\today}

\begin{abstract}
The success of a ferroelectric tunnel junction (FTJ) depends on the asymmetry of electron tunneling as given by the tunneling electroresistance (TER) effect. This characteristic is mainly assessed considering three transport mechanisms: direct tunneling, thermionic emission, and Fowler-Nordheim tunneling. Here, by analyzing the effect of temperature on TER, we show that taking into account only these mechanisms may not be enough in order to fully characterize the performance of FTJ devices. We approach the electron tunneling in FTJ with the non-equilibrium Green function (NEGF) method, which is able to overcome the limitations affecting the three mechanisms mentioned above. We bring evidence that the performance of FTJs is also affected by temperature, in a non-trivial way, via resonance (Gamow-Siegert) states, which are present in the electron transmission probability and are usually situated above the barrier. Although the NEGF technique does not provide direct access to the wavefunctions, we show that, for single-band transport, one can find the wavefunction at any given energy and in particular at resonant energies in the system.
\end{abstract}


\maketitle

\section{Introduction}
\begin{center}
	\begin{figure}[htbp]
		\centerline{\includegraphics[width=2.5in,height=2.5in]{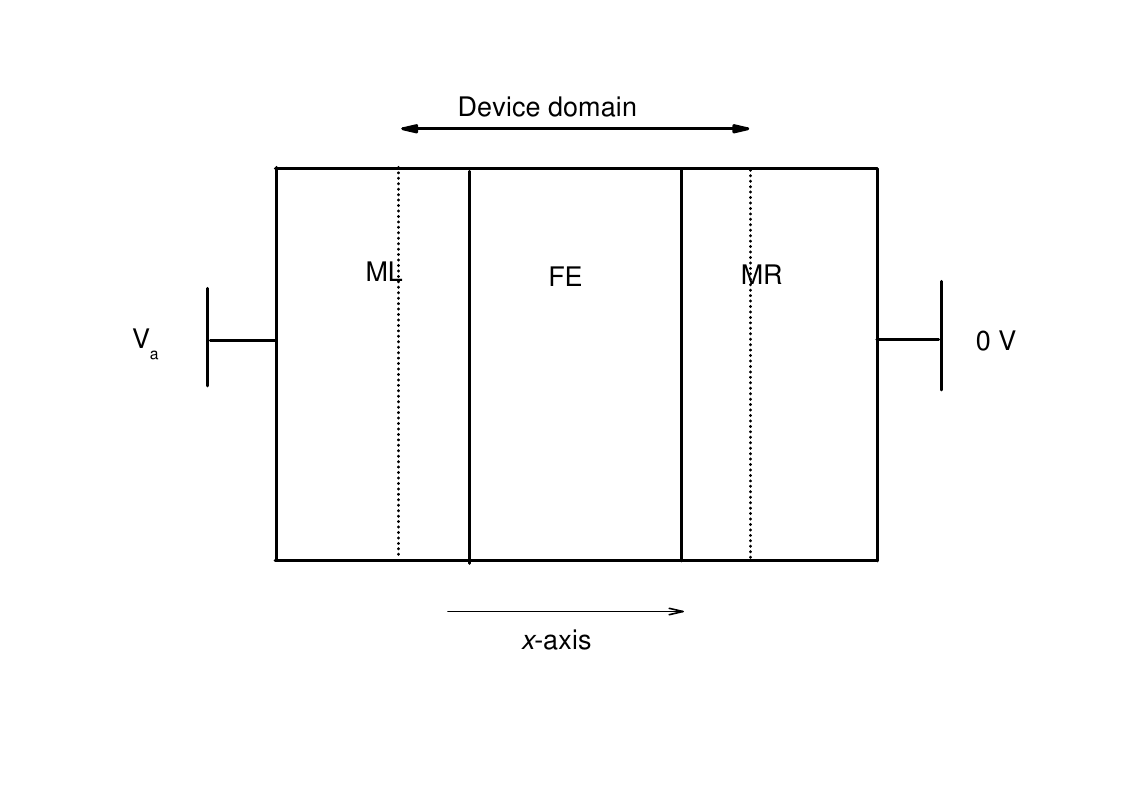}}
		\caption{Schematic representation of an FTJ: a ferroelectric (FE) barrier 
			between two metallic electrodes (ML and MR). The dotted line delimitate the electronic device that is considered for calculation.  } 
		\label{fig1}
	\end{figure}
	
\end{center}

The asymmetric ferroelectric tunnel junction (FTJ), a thin ferroelectric (FE) film 
sandwiched between two dissimilar metallic electrodes or with different interfaces in the
case of the same electrode material(Fig.~\ref{fig1}), is a 
promising electron device for many applications such as low power memories 
\cite{Garcia2014} or neuromorphic computing \cite{Guo2020}. The salient mechanism of FTJ is based on 
the tunneling electroresistance (TER) effect, i. e., a change in the electrical 
resistivity when the electric polarization is reversed under external 
electric field \cite{Zhuravlev2005,Kohlstedt2005}. In other words, the electrical resistance of FTJ 
switches from a high conduction (ON) state to the low conduction (OFF) state 
or vice-versa when a high voltage pulse is applied. Ferroelectrics like 
BaTiO$_{3}$, PbZr$_{0.2}$Ti$_{0.8}$O$_{3}$, BiFeO$_{3}$ \cite{Zenkevich2013,Pantel2012,Hambe2010} as well 
as high-k dielectrics like HfO$_{2}$, Hf$_{0.5}$Zr$_{0.5}$O$_{2}$ \cite{Mueller2012,Mueller2012b} 
have been shown to work as FTJs. The latter are particularly attractive since they a 
compatible with CMOS technology.

The TER effect that is based modulation of the barrier potential upon the 
polarization reversing can be characterized the TER ratio

\begin{equation}
	\label{eq0}
	TER = {\left( {J_{ON} - J_{OFF} } \right)} \mathord{\left/ {\vphantom 
			{{\left( {J_{ON} - J_{OFF} } \right)} {J_{ON} }}} \right. 
		\kern-\nulldelimiterspace} {J_{ON} },
\end{equation}

\noindent
where $J_{ON/OFF}$ are current densities in the two different states. The higher the TER ratio and the $J_{ON}$, the better is
the FTJ performance. The extent of barrier modulation depends on 
numerous factors, such as: i) the thickness and spontaneous polarization of the 
ferroelectric film, ii) the bias voltage, iii) the difference between work functions and also between
screening 
lengths of electrodes, iv) the built-in field and screening 
of the polarization charge, as well as the variation of barrier thickness due 
to piezoelectricity \cite{Wen2020}. 
	
To understand, describe, and model the TER effect, the \textit{ab-intio} methods may provide
  an accurate image of the physics governing FTJ systems like 
the geometry, polarization, electronic structure \cite{Junquera2008},
and transport 
properties \cite{Velev2016,Bragato2018}. Nevertheless,\textit{ab-initio} methods require large 
computational resources, which may prevent their intensive use in the early stages of FTJ design,
when there is a requirement of a quite fast 
scan in parameter space to obtain a device with required functionalities. In the search of optimized devices, 
the semi-empirical methods are fast and, despite all simplifications, are 
extremely useful for the treatment of the charge transport as long as they 
are fed with appropriate physical parameters \cite{Chang2017,Datta2007}. These semi-empirical 
methods are based on the non-equilibrium Green function (NEGF) method that 
can calculate exactly, in principles, the tunneling current and the I-V 
	characteristic in an FTJ \cite{Datta2007}. 

In semi-empirical methods based on NEGF the parameters can be adjusted to 
match the experimental data \cite{Chang2017}. In many cases, in 
order to characterize the electron transport in FTJs and interpret the I-V 
curves, various models of transport mechanisms involving direct tunneling,
thermionic emission, and Fowler-Nordheim tunneling are used \cite{Pantel2010}. 
Direct tunneling presumes low energy for incident electrons such that 
tunneling probability doesn't depend on the profile of the barrier, which is 
assumed to be rectangular on average in the case of Simmons formula \cite{Simmons1963,Gruverman2009} 
or trapezoidal in the case of Brinkman et al. formula \cite{Gruverman2009,Brinkman1970}. Thermionic 
emission is that part of the current carried over the top of the barrier by 
thermally activated charge carriers \cite{Sze2007}. The last mechanism, the 
Fowler-Nordheim tunneling, is used for triangular barrier profiles which are 
encountered at finite applied bias voltages \cite{Fowler1928}. The treatment of the above 
mechanisms is based on some approximations for electron transport. Simmons 
and Brinkman et al. as well as the Fowler-Nordheim formulae are based on the 
semi-classical WKB approximation \cite{Simmons1963,Fowler1928}, while thermionic emission tacitly 
assumes a homogeneous transmission probability of one for all electrons with 
energy above the top of the barrier \cite{Sze2007}. 

In this work, by the exact treatment of the tunneling within the NEGF 
method, we show that the above models describing direct tunneling, 
thermionic emission, and Fowler-Nordheim tunneling are not accurate enough for 
performance characterization of the FTJs. These models cannot capture 
resonant features emerging in FTJ structures that might play an important 
role in carrier transport and ultimately in the performances of device.
The electric current of an impinging electron at a given 
energy is determined by the electron transmission probability at that energy 
weighted (multiplied) by its Fermi-Dirac distribution function. Hence, even 
though they seem unlikely to play a role in the tunneling due to relative 
simple structure of the barriers, the resonances can contribute 
significantly to the electric current since their contribution to 
transmission probability may offset temperature-dependent Fermi-Dirac 
factor. In the following we will show that the resonances emerging close or 
above the top of the barriers play a decisive role for transport at room 
temperature, fact that is not captured by the simple models previously mentioned.
Moreover, resonances may appear in composite (both a ferroelectric and a dielectric)
barriers. The NEGF method 
can accurately describe also these types of heterostructures. 
Generally, the NEGF approach cannot provide full access to the wavefunctions 
especially for multi-band transport \cite{Lake1997}, however for a single-band 
transport we show that this is not the case, the eigenvectors of the 
spectral functions are just the wavefunctions of the system. At this stage 
we are able to identify the Green function of the device as the discrete 
version of the outgoing Green function which is used in diverse scattering 
problems like nuclear reactions \cite{Calderon2010} or electron transport in nanostructures 
\cite{Calderon2010,Calderon1993}. The advantage of such correspondence is that the NEGF of the device 
and the transmission probability can be expanded as a sum of resonance or 
Gamow-Siegert states \cite{Calderon2010,Calderon1993,Tolstikhin1997,Hatano2011}. Thus we are able to sort out the resonances 
that count for the electron transport in FTJs. 

The paper is organized as follows. The next section deals with the 
theoretical background: the calculation of polarization induced 
electrostatic potential across the FTJ, the calculation of transport 
quantities like I-V curve and conductance by NEGF method, and the evaluation 
of wavefunctions and resonance states from NEGF calculations. The third 
section presents numerical results regarding the effect of temperature on 
electric conductance and TER ratio for several practical cases of simple or 
composite BaTiO$_{3}$ based FTJs. In addition, the 
wavefunctions and resonance states for such FTJ structures are analyzed. The fourth 
section summarizes the conclusions of this work. Lastly, in the Appendix we present 
the steps to obtain a discrete tight-binding Hamiltonian from the BenDaniel 
and Duke Hamiltonian \cite{Lake1997}.

\section{Theoretical background}
\subsection{The profile of potential barrier}
In semi-empirical methods the calculation of tunneling current and electric 
conductance is performed by solving simultaneously the electrostatic and 
transport problems They are intricately interrelated since the charge 
density in Poisson equation is calculated self-consistently from the NEGFs 
\cite{Lake1997}. However, when one deals with free carriers only in the metallic 
contacts they can decouple each other. So, let us first deal with the electrostatics and 
the profile of the barrier potential. In the following we treat the 
composite FTJ, where the barrier is composed of a dielectric and a 
ferroelectric layer. If the thickness of dielectric is fixed to $t_{DE}$ and 
that of the ferroelectric to $t_{FE}$, we assume that the dielectric and 
ferroelectric are located for $x$ between -$d_{FE}$ and 0 and for $x$ between 0 and 
$t_{FE}$ , respectively, while the metallic contact ML is at $x <- d_{FE}$ and 
the metallic contact MR is at $x>t_{FE}$ . A widely used and good 
approximation of the electrostatics in metals is the Thomas-Fermi 
approximation \cite{Chang2017,Chen1989}. Within this framework, the electrical fields in 
ML and MR are given by

\begin{equation}
	\label{eq1}
	E_{ML} \left( x \right) = {\tau _S e^{{\left( {x + t_{DE} } \right)} 
			\mathord{\left/ {\vphantom {{\left( {x + t_{DE} } \right)} {\lambda _1 }}} 
				\right. \kern-\nulldelimiterspace} {\lambda _1 }}} \mathord{\left/ 
		{\vphantom {{\tau _S e^{{\left( {x + t_{DE} } \right)} \mathord{\left/ 
							{\vphantom {{\left( {x + t_{DE} } \right)} {\lambda _1 }}} \right. 
							\kern-\nulldelimiterspace} {\lambda _1 }}} {\left( {\varepsilon _1 
						\varepsilon _0 } \right)}}} \right. \kern-\nulldelimiterspace} {\left( 
		{\varepsilon _1 \varepsilon _0 } \right)},
\end{equation}

\begin{equation}
	\label{eq2}
	E_{MR} \left( x \right) = {\tau _S e^{ - {\left( {x - t_{FE} } \right)} 
			\mathord{\left/ {\vphantom {{\left( {x - t_{FE} } \right)} {\lambda _2 }}} 
				\right. \kern-\nulldelimiterspace} {\lambda _2 }}} \mathord{\left/ 
		{\vphantom {{\tau _S e^{ - {\left( {x - t_{FE} } \right)} \mathord{\left/ 
							{\vphantom {{\left( {x - t_{FE} } \right)} {\lambda _2 }}} \right. 
							\kern-\nulldelimiterspace} {\lambda _2 }}} {\left( {\varepsilon _2 
						\varepsilon _0 } \right)}}} \right. \kern-\nulldelimiterspace} {\left( 
		{\varepsilon _2 \varepsilon _0 } \right)}.
\end{equation}

In Eqs (\ref{eq1}) and (\ref{eq2}) \textit{$\tau $}$_{S}$ is the screening charge at ML/dielectric and 
ferroelectric/MR interfaces, \textit{$\lambda $}$_{1}$(\textit{$\varepsilon $}$_{1})$ and \textit{$\lambda $}$_{2}$(\textit{$\varepsilon $}$_{2})$ are the 
Thomas-Fermi screening lengths (dielectric constants) of ML and MR, 
respectively, and \textit{$\varepsilon $}$_{0 }$ is the vacuum permittivity. In the following we 
respectively denote by \textit{$\varepsilon $}$_{FE }$, $E_{FE}$, and $P$ the dielectric constant, the 
electric field, and the intrinsic polarization of ferroelectric and by 
$E_{DE}$, \textit{$\varepsilon $}$_{DE}$ the electric field and the dielectric constant of the 
dielectric. The electrostatic equations are in fact the continuity of the 
normal component of the electric induction $D$ at both ferroelectric/MR and 
ML/dielectric interfaces

\begin{equation}
	\label{eq3}
	\tau _S = \varepsilon _{FE} \varepsilon _0 E_{FE} + P = \varepsilon _{DE} 
	\varepsilon _0 E_{DE} .
\end{equation}

\noindent
Additionally, the bias voltage across the structure obeys the equation

\begin{equation}
	\label{eq4}
	\frac{\tau _S \lambda _1 }{\varepsilon _1 \varepsilon _0 } + \frac{\tau _S 
		\lambda _2 }{\varepsilon _2 \varepsilon _0 } + E_{FE} t_{FE} + E_{DE} t_{DE} 
	+ V + V_{BI} = 0
\end{equation}

\noindent
where $V$ is the applied voltage and $V_{BI}$ is the built-in voltage bias due to 
mismatch of the conduction bands (conduction band discontinuities) and of 
Fermi energies $E_{FL}$ ($E_{FR}$ ) of ML(MR)

\begin{equation}
	\label{eq5}
	V_{BI} = {\left( {\varphi _2 + \varphi _C - E{ }_{FR} - \varphi _1 + E_{FL} } 
		\right)} \mathord{\left/ {\vphantom {{\left( {\varphi _2 + \varphi _C - E{ 
						}_{FR} - \varphi _1 + E_{FL} } \right)} e}} \right. 
		\kern-\nulldelimiterspace} e.
\end{equation}

\noindent
In Eq. (\ref{eq5}) $e$ is the elementary electric charge, \textit{$\phi $}$_{1}$ is the band discontinuity 
at the first interface between ML and the dielectric, \textit{$\phi $}$_{2}$ is the band 
discontinuity at the second interface between the ferroelectric and MR, and 
\textit{$\phi $}$_{C}$ is the band discontinuity at the interface between the dielectric and 
ferroelectric when the ferroelectric is unpolarized. Eliminating $E_{FE}$ and 
$E_{DE}$ from Eqs. (\ref{eq3}) and (\ref{eq4}) we obtain the screening charge \textit{$\tau $}$_{S}$

\begin{equation}
	\label{eq6}
	\tau _S = \frac{V + V_B + {t_{DF} P} \mathord{\left/ {\vphantom {{t_{DF} P} 
					{\left( {\varepsilon _0 \varepsilon _{FE} } \right)}}} \right. 
			\kern-\nulldelimiterspace} {\left( {\varepsilon _0 \varepsilon _{FE} } 
			\right)}}{\frac{\lambda _1 }{\varepsilon _0 \varepsilon _1 } + \frac{\lambda 
			_2 }{\varepsilon _0 \varepsilon _2 } + \frac{t_{DF} }{\varepsilon _0 
			\varepsilon _{FE} } + \frac{t_{DE} }{\varepsilon _0 \varepsilon _{DE} }}.
\end{equation}

\noindent
Now it is straightforward to obtain the electric 
potential and the barrier profile knowing that the electric field is homogeneous in both the dielectric and 
the ferroelectric

\begin{equation}
	\label{eq7}
	U\left( x \right) = \left\{ {\begin{array}{l}
			- e\frac{\tau _s \lambda _1 }{\varepsilon _1 \varepsilon _0 }\exp \left[ {x 
				\mathord{\left/ {\vphantom {x {\lambda _1 }}} \right. 
					\kern-\nulldelimiterspace} {\lambda _1 }} \right],\;x < - t_{DE} \\ 
			- e\tau _s \left( {\frac{\lambda _1 }{\varepsilon _1 \varepsilon _0 } + 
				\frac{t_{DE} }{\varepsilon _{DE} \varepsilon _0 } + \frac{x}{\varepsilon 
					_{DE} \varepsilon _0 }} \right) - \varphi _1 ,\; - t_{DE} \le x < 0 \\ 
			- e\tau _s \left( {\frac{\lambda _1 }{\varepsilon _1 \varepsilon _0 } + 
				\frac{t_{DE} }{\varepsilon _{DE} \varepsilon _0 }} \right) - e\frac{\tau _s 
				- P}{\varepsilon _{FE} \varepsilon _0 }x - \varphi _1 + \varphi _C ,\;0 \le 
			x \le t_{FE} \\ 
			e\left( {V + V_B + \frac{\tau _s \lambda _2 }{\varepsilon _2 \varepsilon _0 
				}\exp \left[ { - {\left( {x - t_{FE} } \right)} \mathord{\left/ {\vphantom 
							{{\left( {x - t_{FE} } \right)} {\lambda _2 }}} \right. 
						\kern-\nulldelimiterspace} {\lambda _2 }} \right]} \right) \\ 
			- \varphi _1 + \varphi _C + \varphi _2 ,{\kern 1pt} {\kern 1pt} x > t_{FE} 
			\\ 
	\end{array}} \right.
\end{equation}

\noindent
In Eq. (\ref{eq7}) $P$ is considered positive when points from ML to MR and negative when it points the other way around.

\subsection{Transport by NEGF}
Transport properties like electric current density and conductance can be 
calculated by solving the Schr\"{o}dinger equation with scattering boundary 
conditions, i. e., either an incoming wave from the left or from the right. 
For a single band, the equation is merely 1D with a BenDaniel and Duke 
Hamiltonian, where the effective mass of electrons can vary across the 
structure \cite{Lake1997}:

\begin{equation}
	\label{eq8}
	H = - \frac{\hbar ^2}{2}\frac{d}{dx}\left( {\frac{1}{m^\ast \left( x 
			\right)}\frac{d}{dx}} \right) + \frac{\hbar ^2k^2}{2m^\ast \left( x \right)} 
	+ U\left( x \right).
\end{equation}

In Eq. (\ref{eq8}) it is assumed that the energy band is parabolic with an isotropic 
effective mass in each layer of the nanostructure, $k$ is the transverse 
wavevector (parallel to each interface), $m^{\ast }$ is the effective mass, 
and $U(x)$ potential energy given by Eq. (\ref{eq7}). Equation (\ref{eq8}) can be discretized into 
1D problem in a tight-binding representation \cite{Lake1997}. The details are presented 
in the Appendix. The Hamiltonian in the matrix format has the following 
tridiagonal form

\begin{equation}
	\label{eq9}
	H = \left( {{\begin{array}{*{20}c}
				{H_L } \hfill & {V_{LD} } \hfill & 0 \hfill \\
				{V_{LD}^\dag } \hfill & {H_D } \hfill & {V_{RD}^\dag } \hfill \\
				0 \hfill & {V_{RD} } \hfill & {H_R } \hfill \\
	\end{array} }} \right).
\end{equation}

\noindent
In this format one can easily see that the Hamiltonian has three parts: the 
semi-infinite left (L) and right (R) metallic electrodes and the device (D) 
that is defined by $H_{D}$, an $n_{D} \quad \times  \quad n_{D}$ matrix. Both 
electrodes act as reservoirs, hence they have well defined chemical 
potentials and temperatures. 
We can define the retarded Green function ($G^{R})$ of the system at energy 
$E$ as the inverse matrix of [($E$ + i\textit{$\eta $}) - $H$], where \textit{$\eta $} = 0+. Similarly the 
advanced Green function $G^{A}$ is the inverse of [(E - i$\eta )-H$], i. 
e.,

\begin{equation}
	\label{eq10}
	G^{R,A}\left( E \right) = \left[ {\mbox{(}E\mbox{ }\pm \mbox{ i}\eta \mbox{) 
			- H}} \right]^{ - 1} .
\end{equation}

\noindent
In the NEGF method one eliminates the degrees of freedom of contacts by 
introducing self-energies into the projected Green functions of the device 
(D)

\begin{equation}
	\label{eq11}
	G_D^{R,A} \left( E \right) = \left[ {\left( {E \pm i\eta } 
		\right) - H_D - \Sigma _B^{R,A} \left( E \right)} \right]^{ - 
		1}.
\end{equation}

\noindent
The self-energy $\Sigma _B^{R,A} \left( E \right) = \Sigma _L^{R,A} \left( E 
\right) + \Sigma _R^{R,A} \left( E \right)$ has two components due to the 
coupling to the left and right contact. It replaces the boundary conditions 
that otherwise would be fulfilled by the construction of a Green function in 
the device region. The self-energy due to the left contact has the following 
expression:

\begin{equation}
	\label{eq12}
	\Sigma _L^{R,A} \left( E \right) = V_{LD} g_L^{R,A} \left( E 
	\right)V_{LD}^\dag ,
\end{equation}

\noindent
where $g_L^{R,A} \left( E \right) = \left[ {\mbox{(}E\mbox{ }\pm \mbox{ 
		i}\eta \mbox{) - H}_L } \right]^{ - 1}$ is the Green function of the 
semi-infinite left contact. Also, $\Sigma _R^{R,A} \left( E \right)$ has a 
similar expression. Here, we notice that due to the fact that we deal with a 
nearest-neighbor tight-binding Hamiltonian, the self-energies $\Sigma _L^R $ 
and $\Sigma _R^R $ are $n_{D} \quad \times  \quad n_{D}$ matrices with just one 
non-zero element, the (1,1) element for $\Sigma _L^R $ and 
($n_{D}$,$n_{D})$ for $\Sigma _R^R $. Thus the retarded Green function is

\begin{equation}
	\label{eq13}
	G_D^R = \left( {{\begin{array}{*{20}c}
				{E\mbox{ + i}\eta - D_1 + \sigma _L^R \left( E \right)} \hfill & { - t_{12} 
				} \hfill & 0 \hfill & \cdots \hfill & 0 \hfill \\
				{ - t_{21} } \hfill & {E\mbox{ + i}\eta - D_2 } \hfill & \ddots \hfill & 
				\ddots \hfill & \vdots \hfill \\
				0 \hfill & \ddots \hfill & \ddots \hfill & { - t_{n_D - 2,n_D - 1} } \hfill 
				& 0 \hfill \\
				\vdots \hfill & \ddots \hfill & { - t_{n_D - 1,n_D - 2} } \hfill & {E\mbox{ 
						+ i}\eta - D_{n_D - 1} } \hfill & { - t_{n_D - 1,n_D } } \hfill \\
				0 \hfill & \cdots \hfill & 0 \hfill & { - t_{n_D ,n_D - 1} } \hfill & 
				{E\mbox{ + i}\eta - D_{n_D } + \sigma _R^R \left( E \right)} \hfill \\
	\end{array} }} \right)^{ - 1}
\end{equation}

The poles of the device Green function are no longer real, an attribute of 
an open quantum system. Starting from Eq. (\ref{eq12}), the calculation of $\sigma 
_L^R \left( E \right)$ involves the calculation of matrix element (0,0) of 
$g_L^R \left( E \right)$, which is the surface Green function of the left 
contact. Bearing in mind that $D_{L}$ and $t_{L}$ are the tight-binding 
parameters of the left contact and defining a longitudinal wavevector 
$k_{l}$, the energy can be parameterized according to the single-band 
dispersion relation for the left contact $E = D_L - 2t_L \cos \left( {k_l 
	\Delta } \right)$. One can find that the matrix element (0,0) of $g_L^R 
\left( E \right)$ is $ - {e^{ik_l \Delta }} \mathord{\left/ {\vphantom 
		{{e^{ik_l \Delta }} {t_L }}} \right. \kern-\nulldelimiterspace} {t_L }$ and 
$\sigma _L^R \left( E \right) = - t_L e^{ik_l \Delta }$ \cite{Chang2017,Lake1997,Chen1989}. The 
expression of $g_L^R \left( E \right)$ is calculated with outgoing boundary 
condition \cite{Chen1989}, hence the self-energy is for this kind of boundary 
condition. 

One can further define: (a) the spectral function $A = i\left( {G^r - G^a} 
\right)$, which also has a matrix form, whose diagonal is just local density 
of states up the 2$\pi $ factor, and (b) the broadening function due to the 
coupling to the left and right contacts $\Gamma _{L,R} \left( E \right) = 
i\left( {\Sigma _{L,R} \left( E \right) - \Sigma _{L,R}^\dag \left( E 
	\right)} \right)$. Using the projection operator on the device $D$ one can show 
that the projection of the full spectral function $A$ on the device space is 
\cite{Paulsson2007}



\begin{equation}
	\label{eq14}
	A{ }_D\left( E \right) = i\left( {G_D^R \left( E \right) - G_D^A \left( E 
		\right)} \right) = G_D^R \left( E \right)\left( {\Gamma _L \left( E \right) 
		+ \Gamma _R \left( E \right)} \right)G_D^A \left( E \right).
\end{equation}

Moreover one can further show that partial spectral densities

\begin{equation}
	\label{eq15}
	A_{L,R} \left( E \right) = G_D^R \left( E \right)\Gamma _{L,R} \left( E 
	\right)G_D^A \left( E \right)
\end{equation}

\noindent
are spectral densities due to incident Bloch waves that come respectively 
from the left (L) and from right (R) \cite{Paulsson2007}. Thus $G_D^R \left( E \right)$ 
contains information about both solutions of the scattering problem. 
Furthermore it can be shown that the current flow from an incident Bloch 
wave that comes from the left electrode into the right electrode is

\begin{equation}
	\label{eq16}
	j\left( E \right) = \frac{2e}{h}Tr\left[ {G_D^R \left( E \right)\Gamma _L 
		\left( E \right)G_D^A \left( E \right)\Gamma _R \left( E \right)} \right]
\end{equation}

\begin{center}
	\begin{figure}[htbp]
		\centerline{\includegraphics[width=2.60in,height=1.74in]{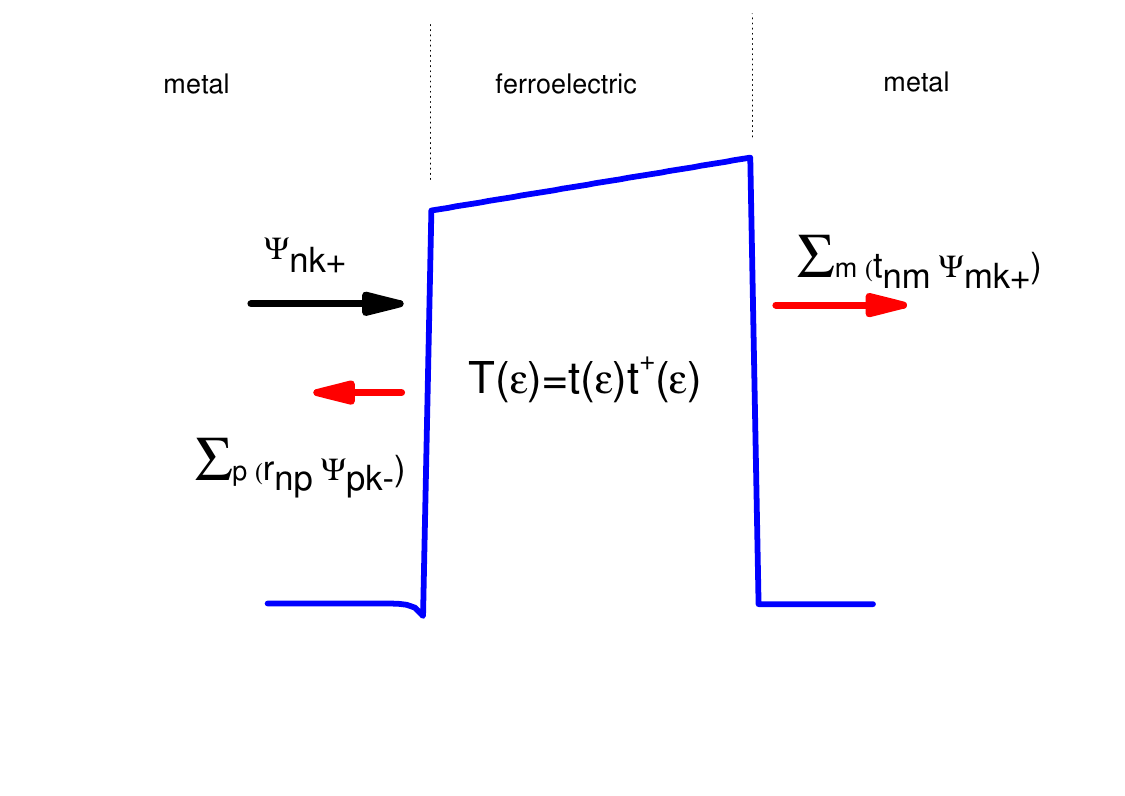}}
        \caption{The behavior of an incident Bloch wave $\Psi _{nk + }$coming from 
	the left. A part is reflected with the coefficient $r$ and the other part is 
	transmitted with a coefficient $t.$ In the case of a single-band transport the 
	coefficients $r$ and $t$ are simple scalars.}
        \label{fig2}
	\end{figure}
	
\end{center}

\noindent
$G_D^R \left( E \right)\Gamma _L \left( E \right)G_D^A \left( E \right)\Gamma 
_R \left( E \right)$ is just the matrix of transmission probability $T(E)$
that has a schematic representation in Fig.~\ref{fig2}. From Eq. (\ref{eq16}) the 
expression of the total current takes the form of Landauer-B\"{u}ttiker 
formula \cite{Datta2007,Lake1997}:

\begin{equation}
	\label{eq17}
	J = \frac{2e}{h}\int {Tr\left( {T\left( E \right)} \right)\left( {f_L \left( 
			E \right) - f_R \left( E \right)} \right)dE} .
\end{equation}

In Eq. (\ref{eq17}) the Tr() operation includes both the trace over the $T\left( E 
\right)$ matrix and the integration over the transverse wavevector 
\textit{k} . The functions $f_{L,R}$ are the Fermi-Dirac 
distribution functions of the $L$ and $R$ electrodes. Performing the trace 
operation only over matrix $T$ we obtain the transmission probability coefficient $\tilde{t}$ and the 
the Landauer-B\"{u}ttiker formula becomes

\begin{equation}
	\label{eq18}
	J = \frac{e}{2\pi ^2h}\int\limits_{ - \infty }^\infty {\int\limits_{ - 
			\infty }^\infty {d^2k\int\limits_0^\infty {\tilde{t}\left( {k,E} \right)\left( {f_L 
					\left( E \right) - f_R \left( E \right)} \right)dE} } } . \quad .
\end{equation}

In addition, in the linear regime (small bias voltages $V)$ and at temperature 
of 0K we obtain the Landauer conductance formula \cite{Datta2007}

\begin{equation}
	\label{eq19}
	G = \frac{2e^2}{h}\int\limits_{ - \infty }^\infty {\int\limits_{ - \infty 
		}^\infty {\frac{d^2k_t }{\left( {2\pi } \right)^2}\tilde{t}\left( {k_t ,E_F } 
			\right)} } .
\end{equation}

\subsection{Retrieving the wavefunction from the spectral function. Resonance states }

Let us consider the spectral function $A_L \left( E \right) = G_D^R \left( E 
\right)\Gamma _L \left( E \right)G_D^A \left( E \right)$ . It signifies the 
projected spectral function on the device for incident waves coming from the 
left. As it was pointed out in Ref. \cite{Paulsson2007}, in general the eigenvectors of 
$A_{L}$ cannot be identified with the wavefunction in the device region since 
there are several eigenvectors of $A_{L}$ with non-zero eigenvalues. This is 
rather obvious in the tight-binding representation but for multi-band 
problem. For a single-band problem, however, this is not the case. $A_{L}$ 
has just one non-zero eigenvalue. Its corresponding eigenvector is just the 
function that is proportional to the wavefunction in the device region. This 
statement can be proven directly. The matrix $\Gamma _L \left( E \right)$ 
has just one element different from 0 like its corresponding self-energies 
$\Sigma _L^{R,A} \left( E \right)$

\begin{equation}
	\label{eq20}
	\Gamma _L \left( E \right) = \left( {{\begin{array}{*{20}c}
				{\gamma _L \left( E \right)} \hfill & 0 \hfill & \cdots \hfill \\
				0 \hfill & 0 \hfill & \hfill \\
				\vdots \hfill & \hfill & \ddots \hfill \\
	\end{array} }} \right).
\end{equation}

\noindent
Denoting by $G_{D,i,j} \left( E \right)$ the matrix elements of $G_D^R \left( 
E \right)$ and by $A_{L,i,j} \left( E \right)$ the matrix elements of $A_L 
\left( E \right)$, it is easy to check that $A_{L,i,j} \left( E \right) = 
\gamma _L \left( E \right)G_{D,i,1} \left( E \right)G_{D,j,1}^\ast \left( E 
\right)$, where the * means complex conjugation. One can further see that 
the $n_{D}$-dimensional vector

\begin{equation}
	\label{eq21}
	v_L = \left( {{\begin{array}{*{20}c}
				{G_{D,1,1} \left( E \right)} \hfill & {G_{D,2,1} \left( E \right)} \hfill & 
				{{\begin{array}{*{20}c}
							\cdots \hfill & {G_{D,n_D ,1} \left( E \right)} \hfill \\
				\end{array} }} \hfill \\
	\end{array} }} \right)
\end{equation}

\noindent
is an eigenvector of $A_L \left( E \right)$ with the eigenvalue

\begin{equation}
	\label{eq22}
	\lambda _L = \gamma _L \left( E \right)\sum\limits_{i = 1}^{n_D } {\left| 
		{G_{D,1,i} \left( E \right)} \right|^2} = Tr\left( {A_L \left( E \right)} 
	\right).
\end{equation}

The fact that $\lambda _L = Tr\left( {A_L \left( E \right)} \right)$ ensures 
that \textit{$\lambda $}$_{L}$ is the only non-zero eigenvalue of $A_L \left( E \right)$. In a 
bra and ket notation we thus write $A_L \left( E \right)$ as

\begin{equation}
	\label{eq23}
	A_L \left( E \right) = \gamma _L \left( E \right)\left| {v_L } \right\rangle 
	\left\langle {v_L } \right|
\end{equation}

\noindent
since the norm of $v_L $ is just $\sqrt {{\lambda _L } \mathord{\left/ 
		{\vphantom {{\lambda _L } {\gamma _L }}} \right. \kern-\nulldelimiterspace} 
	{\gamma _L }} $. Eqs. (\ref{eq21}) and (\ref{eq23}) of $v_L $ and $A_L \left( E \right)$ 
guarantee that $v_L $ is proportional to the solution of Schr\"{o}dinger 
equation for incoming waves from the left projected on the device space

\begin{equation}
	\label{eq24}
	\left| {\Psi _L } \right\rangle _D = \sqrt {\frac{\gamma _L }{2\pi }} \,v_L 
\end{equation}

\noindent
Similar calculations can be performed for $A_R \left( E \right)$, 
explicitly, the matrix form is $A_{R,i,j} \left( E \right) = \gamma _R 
\left( E \right)G_{D,i,N} \left( E \right)G_{D,j,N}^\ast \left( E \right)$ 
with the eigenvector

\begin{equation}
	\label{eq25}
	v_R = \left( {{\begin{array}{*{20}c}
				{G_{D,1,n_D } \left( E \right)} \hfill & {G_{D,2,n_D } \left( E \right)} 
				\hfill & {{\begin{array}{*{20}c}
							\cdots \hfill & {G_{D,n_D ,n_D } \left( E \right)} \hfill \\
				\end{array} }} \hfill \\
	\end{array} }} \right)
\end{equation}

\noindent
the eigenvalue

\begin{equation}
	\label{eq26}
	\lambda _R = \gamma _R \left( E \right)\sum\limits_{i = 1}^{n_D } {\left| 
		{G_{D,i,n_D } \left( E \right)} \right|^2} = Tr\left( {A_R \left( E \right)} 
	\right),
\end{equation}

\noindent
and the simple bra and ket form

\begin{equation}
	\label{eq27}
	A_R \left( E \right) = \gamma _R \left( E \right)\left| {v_R } \right\rangle 
	\left\langle {v_R } \right|.
\end{equation}

\noindent
Also, Eqs. (\ref{eq25}) and (\ref{eq27}) of $v_R $ and $A_R \left( E \right)$ guarantee that 
$v_R $ is proportional to the solution of Schr\"{o}dinger equation for 
incoming waves from the right projected on the device space

\begin{equation}
	\label{eq28}
	\left| {\Psi _R } \right\rangle _D = \sqrt {\frac{\gamma _R }{2\pi }} \,v_R .
\end{equation}

\noindent
The total spectral function $A_D \left( E \right)$ is the sum of $A_L \left( 
E \right)$ and $A_R \left( E \right)$, hence its range is spanned by $v_L $ 
and $v_R $ with two eigenvalues \textit{$\lambda $}$_{1}$ and \textit{$\lambda $}$_{2}$. They obey the following 
equation: $\lambda _1 + \lambda _2 = \lambda _L + \lambda _R $. Also, it is 
easy to find that the squared modulus of the overlap between
$\left| {\Psi_L } \right\rangle _D $ and $\left| {\Psi _R } \right\rangle _D $ is

\begin{equation}
	\label{eq29}
	\left| {{ }_D\left\langle {\Psi _L } \right|\left. {\Psi _R } \right\rangle 
		_D } \right|^2 = {\left( {\lambda _L \lambda _R - \lambda _1 \lambda _2 } 
		\right)} \mathord{\left/ {\vphantom {{\left( {\lambda _L \lambda _R - 
						\lambda _1 \lambda _2 } \right)} {4\pi ^2}}} \right. 
		\kern-\nulldelimiterspace} {4\pi ^2}.
\end{equation}

\noindent
From the analysis we are going to perform in the next section on 
realistic examples we will see that for energies in the direct tunneling regime there 
are two distinct solutions with low overlap. In this case \textit{$\lambda $}$_{1}$ and 
\textit{$\lambda $}$_{2}$ are close to \textit{$\lambda $}$_{L}$ and \textit{$\lambda $}$_{R}$ . In the opposite case, at resonance, 
one of the two eigenvalues \textit{$\lambda $}$_{1}$ or \textit{$\lambda $}$_{2}$ is much larger than the other, 
hence the overlap is large. 
In a similar manner we can obtain the explicit expression of the 
transmission probability in terms of the Green function

\begin{equation}
	\label{eq30}
	\tilde{t}\left( E \right) = \gamma _L \left( E \right)\gamma _R \left( E 
	\right)\left| {G_{D,1,n_D } } \right|^2.
\end{equation}

\noindent
Eq. (\ref{eq30}) has been previously deduced using an iterative procedure to 
calculate the Green function \cite{Lake1997,Klimeck1995}. It additionally shows that the 
transmission probability has the spectral properties of the Green function. 

There is a large body of work in which the Green function $G_D^R \left( E 
\right)$ is set in a meaningful representation. Such a representation is 
given by the expansion of the Green function in resonance (Siegert-Gamow) 
states \cite{Calderon2010,Calderon1993}, which lead to Breit-Wigner formula for resonances in 
transmission. Here we will outline a few results about resonant states 
representation that are connected with our results discussed above. The 
expansion of the Green function in resonance states is the sum over these 
states plus a background and it looks like the following

\begin{equation}
	\label{eq31}
	G_D^R \left( {x,x',E} \right) = \sum\limits_{n = 1}^N {\frac{u_n \left( x 
			\right)u_n \left( {x'} \right)}{E - E_n }} + B\left( E \right),
\end{equation}

\noindent
where $u_n \left( x \right)$ is a resonant state that satisfies the 
Schrodinger equation

\begin{equation}
	\label{eq32}
	H_D u_n \left( x \right) = E_n u_n \left( x \right)
\end{equation}

\noindent
with outgoing boundary conditions at the boundary of the device. Since these 
boundary conditions are not hermitian, the eigenvalues are rather complex, i. 
e., $E_n = E_{rn} - i{\Gamma _n } \mathord{\left/ {\vphantom {{\Gamma _n }2}} \right. \kern-\nulldelimiterspace} 2$.
Resonant states come into pairs 
with a negative and positive imaginary part \cite{Calderon2010,Tolstikhin1997,Hatano2011}. Now, it is rather 
obvious that for energy $E$ far from any resonant energy $E_{rn}$ the solution 
of Schr\"{o}dinger equation for incoming waves from the left is different 
from the solution of Schr\"{o}dinger equation for incoming waves from the 
right due to different mixing of the tails of various resonances. However, 
for energy $E$ near a resonant energy $E_{rn}$ both solution from the left and 
from the right are similar since the dominant term is that given by $u_n 
\left( x \right)$. Finally, there is quite straightforward to notice by using 
Eq. (\ref{eq30}) that transmission probability coefficient has a multi-resonance Breit-Wigner 
like form  \cite{Calderon1993}

\begin{equation}
	\label{eq33}
	\tilde{t}\left( E \right) = \sum\limits_{n = 1}^N {T_n \left( E \right)} + 
	\sum\limits_{n < m}^N {T_{nm} \left( E \right)} + R\left( E \right).
\end{equation}

\noindent
The term

\begin{equation}
	\label{eq34}
	T_n \left( E \right) = \frac{A_n }{\left( {E - E_{rn} } \right)^2 + \left( 
		{{\Gamma _n } \mathord{\left/ {\vphantom {{\Gamma _n } 2}} \right. 
				\kern-\nulldelimiterspace} 2} \right)^2}
\end{equation}

\noindent
is the Breit-Wigner expression, $T_{nm} \left( E \right)$ is the interference 
term between resonances, and $R\left( E \right)$ is the term resulting from 
the background.

\section{Numerical analysis of tunneling in relevant FTJs}
\subsection{Temperature influence on conductance and TER ratio}
Numerically we calculated all quantities defined in the previous section 
using NEGF formalism. 
The calculation of I-V characteristics is performed with Eq. (\ref{eq18}), which is 
able to reproduce the non-linear regime for large bias voltages. Equation 
(\ref{eq19}) describe the linear regime at 0K and is often used 
(especially in ab-initio calculations, where a full I-V curve is extremely 
costly) in the calculation of the TER ratio 
defined by Eq.~(\ref{eq0}). In the following we shall 
analyze a few practical FTJ structures.

\begin{center}
\begin{figure}[htbp]
	\centerline{\includegraphics[width=2.50in,height=2.5in]{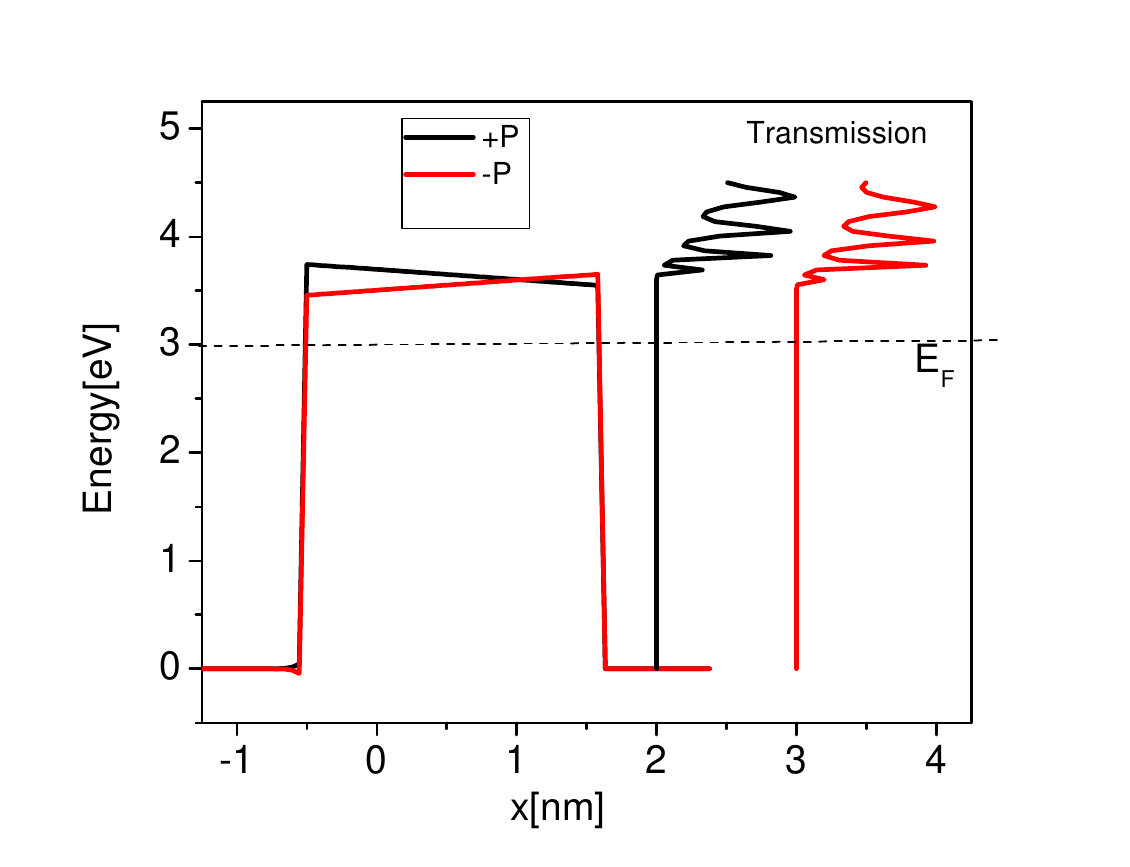}}

	\caption{Potential profiles of a BaTiO$_{3 }$ barrier of 2 nm thickness 
		between Pt and SrRuO$_{3}$ metallic contacts for both directions of 
		polarization. Schematically on the right hand side are 
		shown the transmission probability coefficients.}
        \label{fig3}
\end{figure}
\end{center}

\subsubsection{Pt/BaTiO$_{3}$/SrRuO$_{3}$ FTJ}

\begin{figure}[htp]
	\begin{center}
		\subfigure {\label{fig4a} \includegraphics [width=2.5in,height=2.5in]{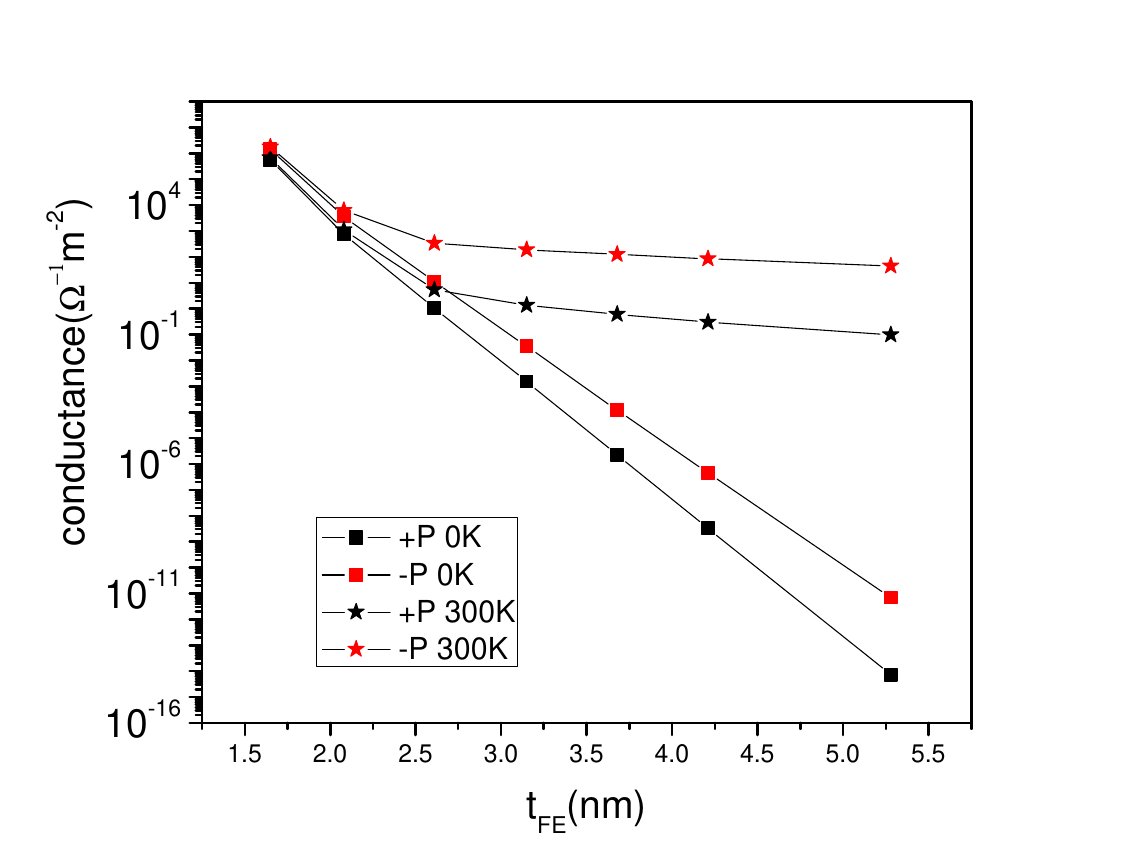} }  
		\subfigure {\label{fig4b} \includegraphics [width=2.50in,height=2.5in]{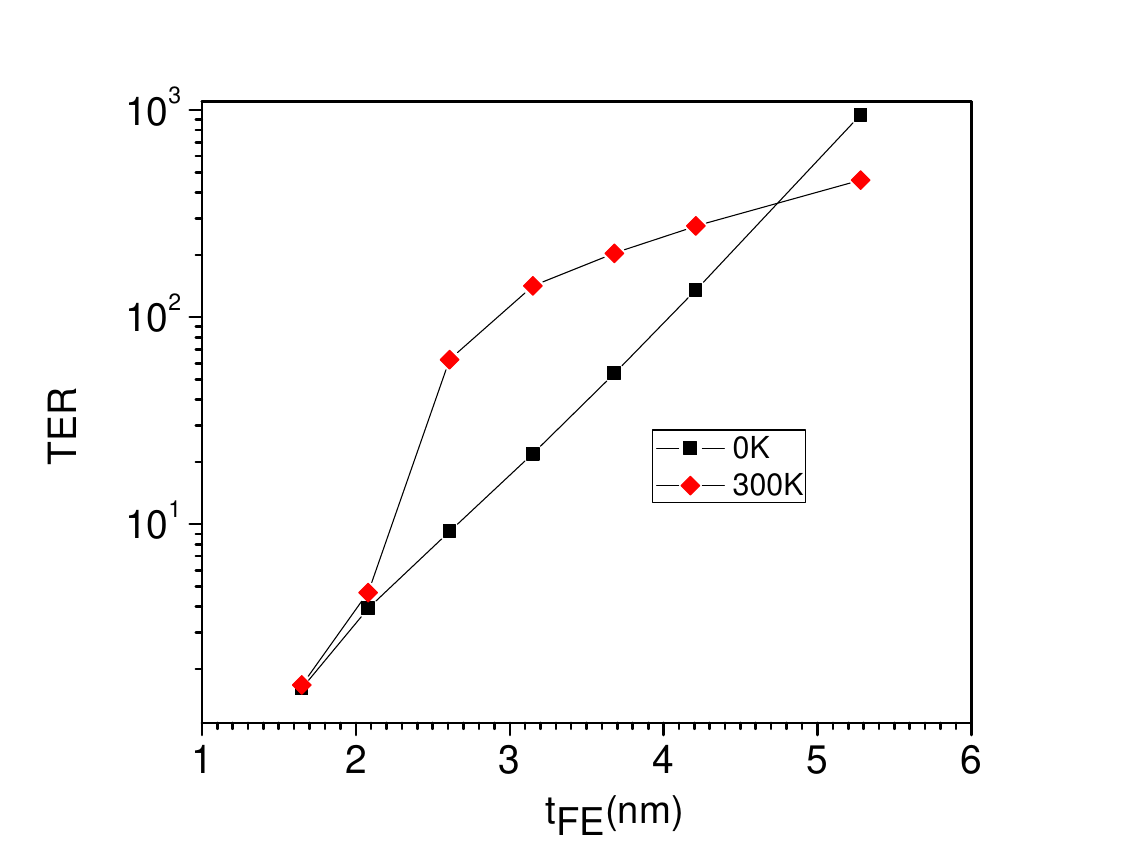} } 
	\end{center}
	\caption{Pt/BaTiO$_{3}$/SrRuO$_{3}$ FTJ:(a) Conductance and (b) TER ratio as a function of the barrier thickness at 0K and 300K.}
	\label{fig4}
\end{figure}

The first system to deal with is that of a BaTiO$_{3}$ ferroelectric barrier on top of a 
SrRuO$_{3}$ substrate which acts as contact layer, and on top of BaTiO$_{3}$ 
is the Pt contact, which is considered as the left contact. Physical 
parameters are: $\lambda _{1} =0.045$ nm, $\lambda_{2} =0.075$ nm, $\varepsilon_{1}=2$, \textit{$\varepsilon $}$_{2}$=8.45, 
\textit{$\varepsilon $}$_{FE}$=125, $E_{FL}=E_{FR}$=3 eV, \textit{$\phi $}$_{1}$=\textit{$\phi $}$_{2}$=3.6 eV, $P$=16 $\mu 
$C/cm$^{2}$. The effective masses are: 5 m$_{0}$ for SrRuO$_{3}$, 2 m$_{0}$ 
for BaTiO$_{3}$, and m$_{0}$ for Pt, where m$_{0}$ is the free electron mass 
\cite{He2019}. The Fermi energy is set to  $E_F = E_{FL}=E_{FR}$=3 eV. One should note
that the electron effective masses are different in the three layers of the structure, hence the 
calculations need full numerical integration over transverse 
\textit{k} wavevector in Eqs. (\ref{eq18}) and (\ref{eq19}). 
In Fig.~\ref{fig3} we show the potential profiles of a 2 nm thick BaTiO$_{3}$ barrier 
for both directions of polarization. The transmission probability coefficients, also depicted
in Fig. \ref{fig3} right, exhibit resonances below and above the very top of the 
barriers. The resonances are associated with the peaks in the transmission 
spectra and if they are well resolved they obey Eq. (\ref{eq34}). 
In Fig.~\ref{fig4a} we plot the conductance of the system and in Fig.\ref{fig4b} the TER 
ratio at 0K and 300K.

\begin{figure}[htp]
	\begin{center}
		\subfigure {\label{fig5a} \includegraphics [width=2.5in,height=2.5in]{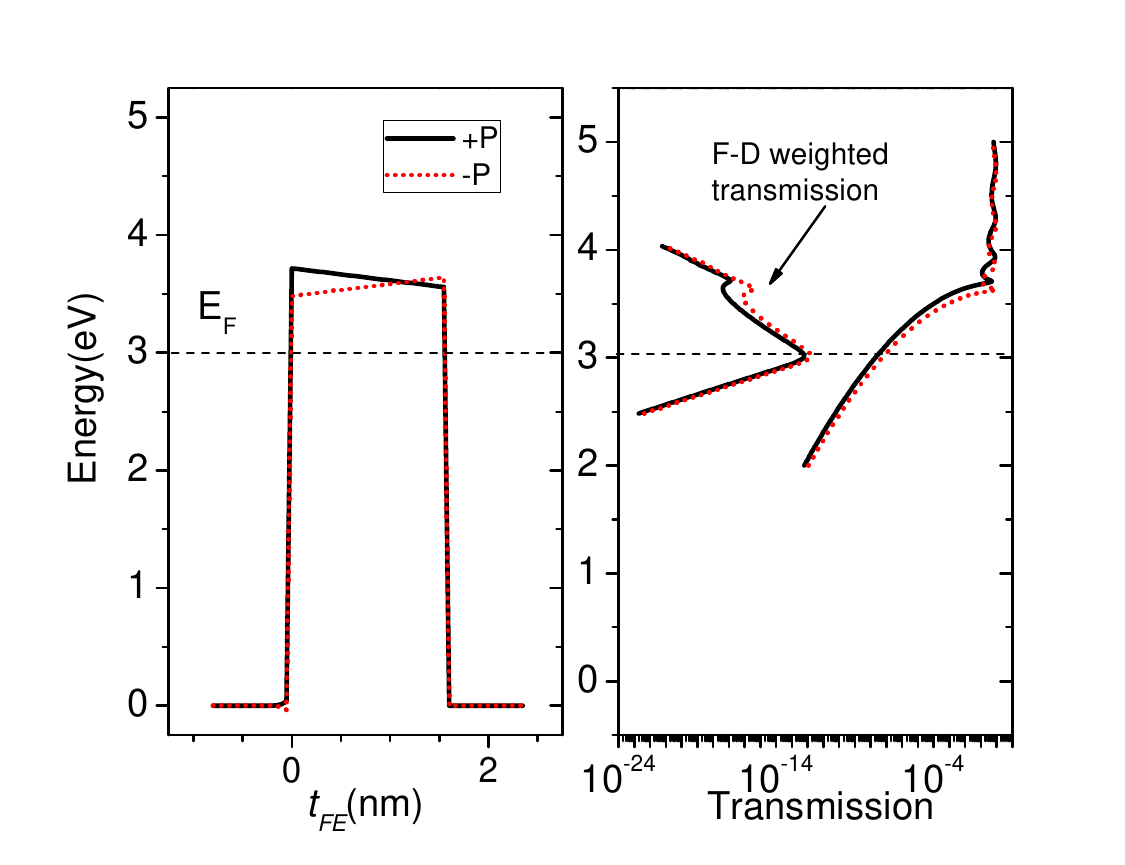}}  
		\subfigure {\label{fig5b} \includegraphics [width=2.50in,height=2.5in]{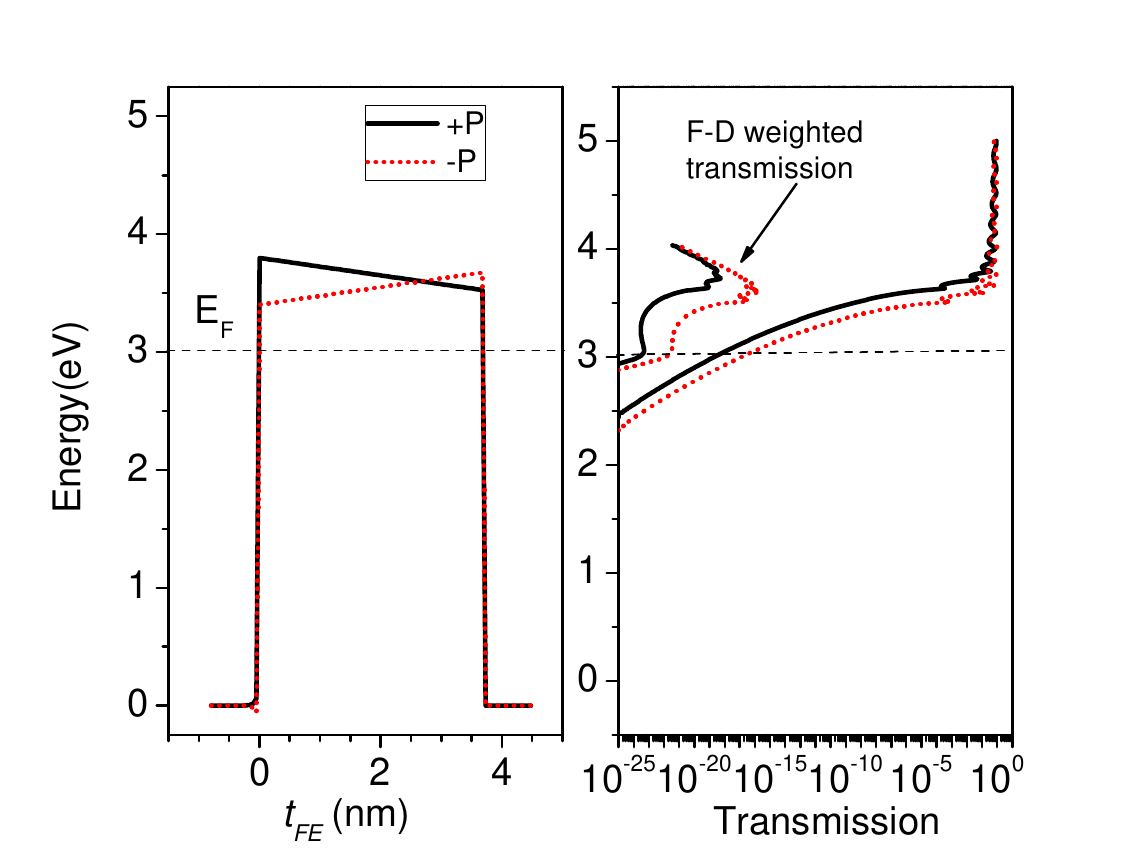}} 
	\end{center}
	\caption{Pt/BaTiO$_{3}$/SrRuO$_{3}$ FTJ: the barrier profile, the 
		transmission probability coefficient, and the Fermi-Dirac weighted transmission at 300K 
		for two barrier thicknesses - (a) 1.6 nm and (b) 3.5 nm. The dashed line in 
		the graphs indicates the Fermi energy.}
	\label{fig5}
\end{figure}

The calculation of the conductance at 300 K was performed with an applied 
bias of 0.0001 V which is low enough to satisfy the linear regime 
conditions. Both the conductance and TER ratio behave 
differently at 300K with respect to 0K. At 300K we observe two regimes 
depending on the ferroelectric thickness, one regime up to 2.5 nm and
another one beyond that value. The difference between the two regimes is 
explained in Fig.~\ref{fig5}. At 0K the channels open to electron 
flow are those up to Fermi energy represented by dashed line. These channels 
are those of direct tunneling such that the Simmons and Brinkman formulae 
are appropriate \cite{Simmons1963,Gruverman2009,Brinkman1970}. For the barrier of 1.6 nm thickness (Fig.~\ref{fig5a}) the 
transport occurs around Fermi energy even at room temperature (300K), hence 
that similar behavior of the conductance at 300K with respect to 0K, see Fig.~\ref{fig4a}. There 
is a small contribution to the transmitted current from the resonance 
states, but the contribution is orders of magnitude smaller. However, for 
thicker barriers, like the one shown in the Fig.~\ref{fig5b}, things may change. 
First, the direct tunneling current is much smaller since it has an 
exponential dependence on thickness. Second, the resonance levels are spaced much 
closer, hence their contribution increases considerably. Even if their
occupancy, given by Fermi-Dirac function, might be small, it 
is offset by the large transmission probability. In Fig.~\ref{fig5b} it can be easily 
seen that the contribution from resonance states is overwhelmingly larger 
than the contributions from states near the Fermi energy. Moreover, electrons
with energies closer to the barrier top encounter a triangular barrier profile
and so they may still reach a resonance state; this mechanism cannot be 
described within the semiclassical WKB theory of Fowler-Nordheim \cite{Fowler1928}. A similar behavior 
of the TER ratio at room temperature was also observed in experimental data 
\cite{Wang2016}. It was found even a degradation of TER when increasing the 
ferroelectric thickness. However, performing their analysis based on Brinkman model at 
finite bias voltage, the authors attributed the TER degradation to the rather high 
levels of noise in the measurements.

\subsubsection{Pt/SrTiO$_3$ /BaTiO$_3$ /SrRuO$_3$ composite barrier FTJ}
\begin{figure}[htp]
	\begin{center}
		\subfigure {\label{fig6a} \includegraphics [width=2.5in,height=2.5in]{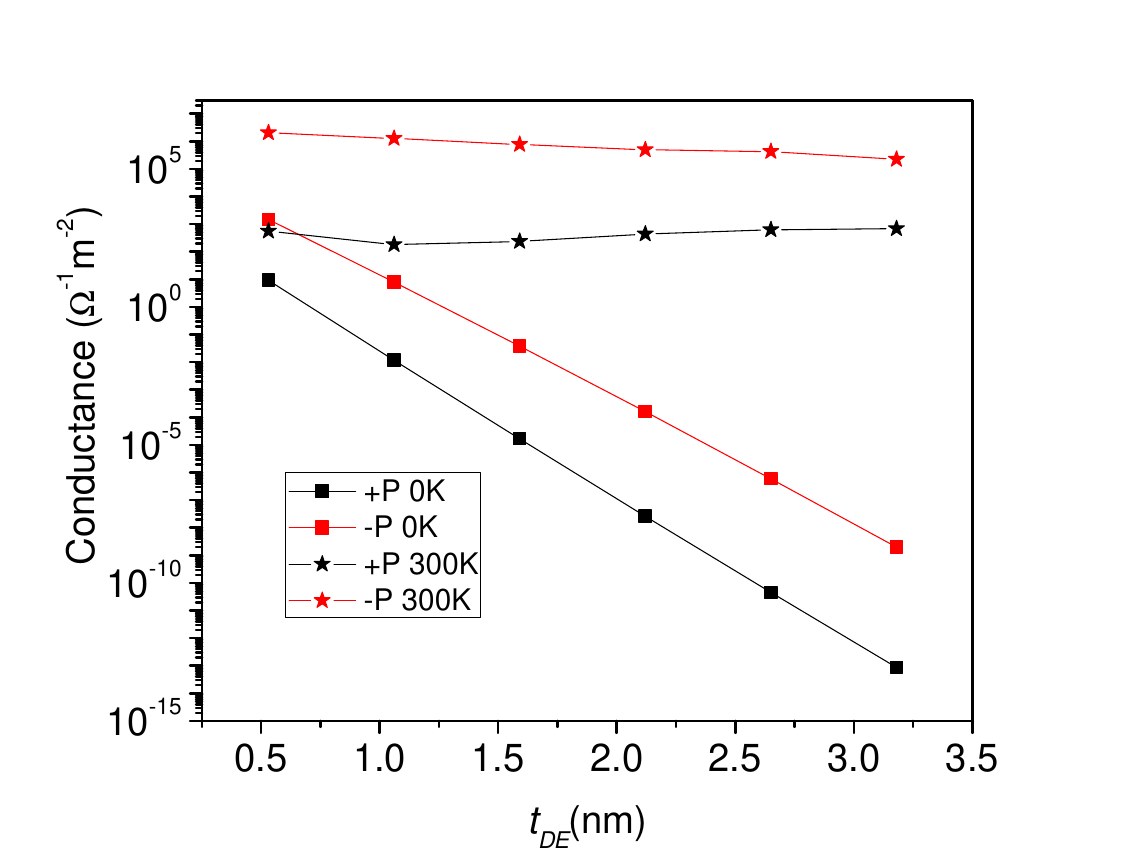}}  
		\subfigure {\label{fig6b} \includegraphics [width=2.50in,height=2.5in]{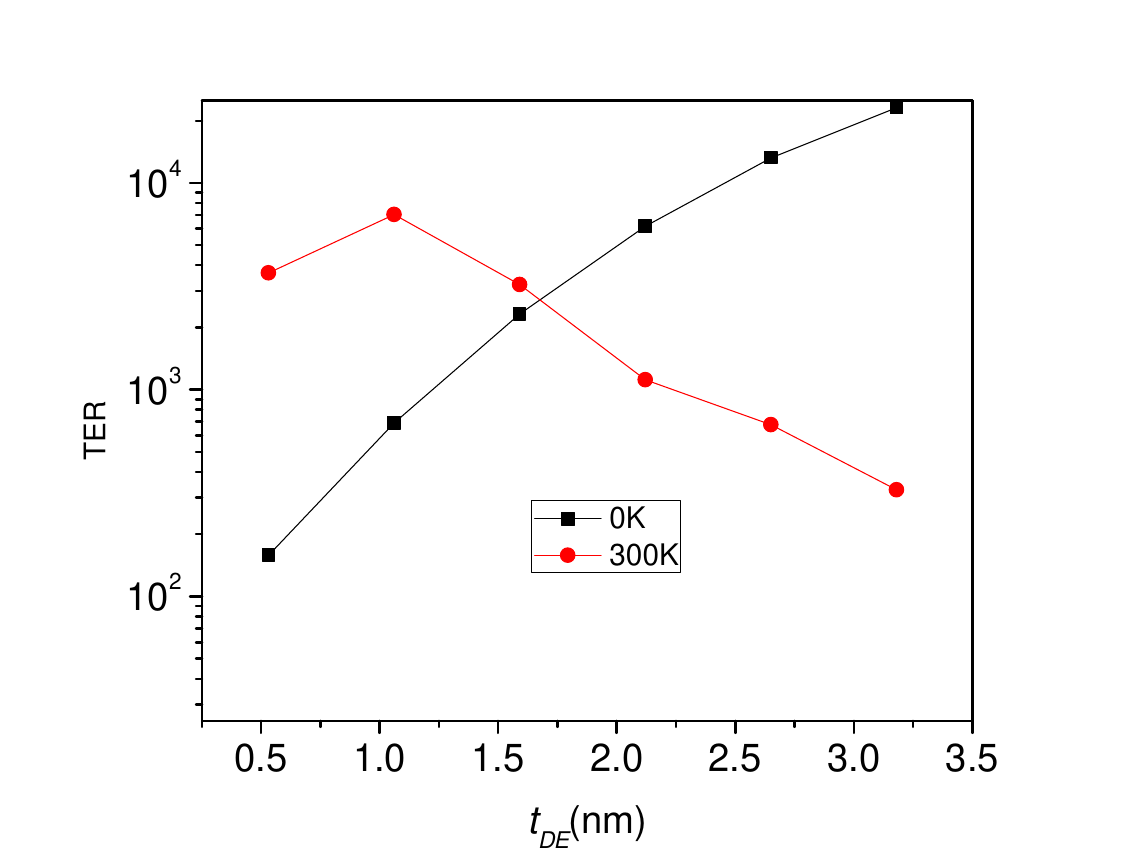}} 
	\end{center}
	\caption{Pt/BaTiO$_{3}$/BaTiO$_{3}$/SrRuO$_{3}$ composite FTJ:(a) Conductance 
		and (b) TER ratio as a function of the dielectric thickness at 0K and 300K.}
	\label{fig6}
\end{figure}

The second system we analyze is a composite TFJ, where a dielectric barrier 
(SrTiO$_{3})$ is added beside the BaTiO$_{3}$ ferroelectric barrier. The 
electrodes are Pt (left) and SrRuO$_{3}$ (right). The role of the dielectric 
layer is to increase the asymmetry of the system, hence it is expected a 
higher TER ratio \cite{Zhuravlev2009}. Physical parameters are slightly changed with respect 
the previous case \cite{Zhuravlev2009,Ma2019}: $\lambda_{1}$=0.045 nm, $\lambda_{2}$=0.08 nm, $\varepsilon_{1}$=2, 
$\varepsilon_{2}$=8.45, $\varepsilon_{FE}$=90, $\varepsilon_{DE}$=90,, $E_{FL}=E_{FR}$=3 eV, $\phi_{1}$=$\phi_{2}$=3.6 
eV, $\phi_{C}$=0 eV, $P$=20 $\mu $C/cm$^{2}$. The effective masses are: 5 m$_{0}$ 
for SrRuO$_{3}$, 2 m$_{0}$ for BaTiO$_{3}$ and SrTiO$_{3}$, and m$_{0}$ for 
Pt. In the following,  BaTiO$_{3}$ thickness is fixed to 2.4 nm and we have varied the SrTiO$_{3}$ thickness
from 0.5 nm to 3 nm. The results of calculations are presented 
in Fig.~\ref{fig6}.

\begin{figure}[htbp]
	\centerline{\includegraphics[width=2.50in,height=2.5in]{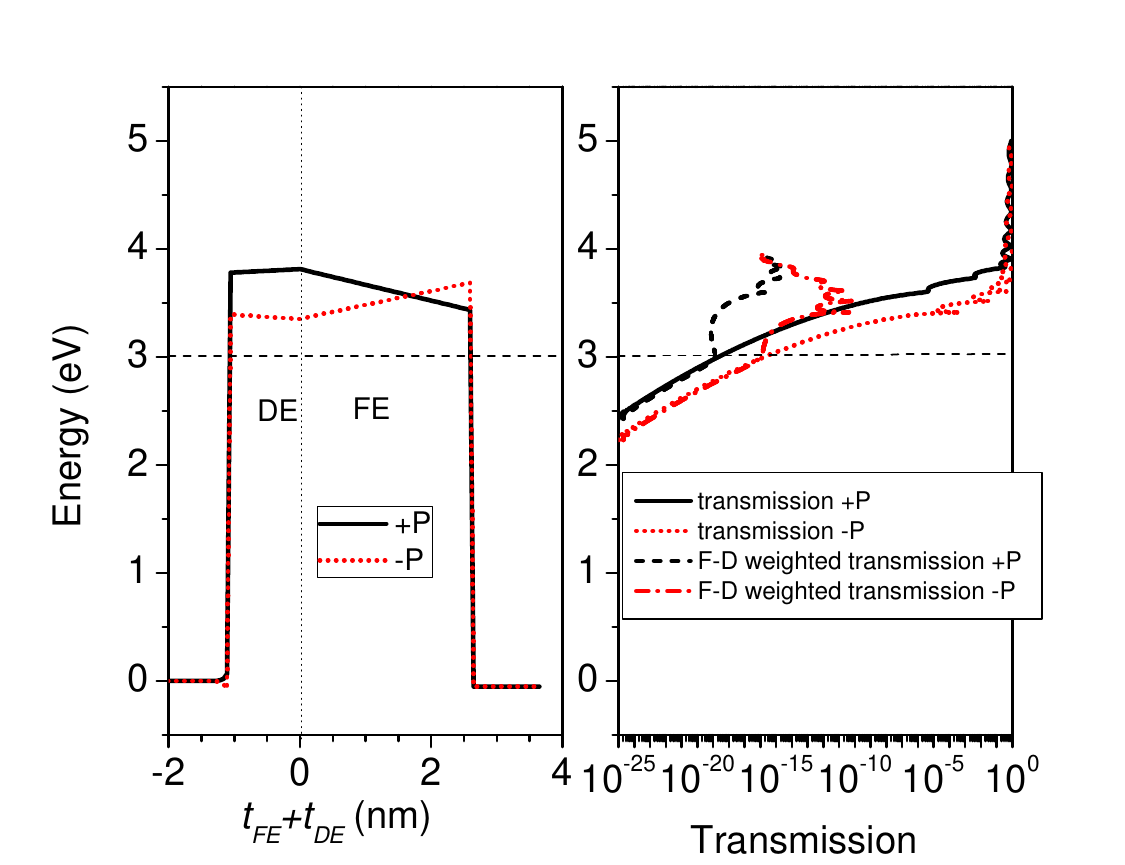}}
		\caption{ Pt/SrTiO$_{3}$/BaTiO$_{3}$/SrRuO$_{3}$ composite FTJ: the barrier profile, the transmission probability coefficient, and the Fermi-Dirac weighted transmission probability at 300K; $t_{FE}$=2.5 nm, $t_{DE}$=1 nm. The dashed line in the graphs indicates the Fermi energy.}
	\label{fig7}
\end{figure}

At 0K the conductance for both polarization directions shows an exponential dependence 
on dielectric thickness. At room temperature, on the other hand, it appears 
that this exponential dependence is no longer valid, as revealed by the TER ratio.
In this case, a decrease of TER ratio takes place when the dielectric 
thickness increases (Fig.~\ref{fig6b}). The explanation of TER degradation is 
provided by the results presented in Fig.~\ref{fig7}. One may observe that the
electric charges are 
mainly transported through quantum states that a close or above the barrier 
top, the contribution of the states near the Fermi energy being negligible . 
The total barrier thickness is 3 nm at least, a value at which the resonance 
states start to play a significant role. As the barrier thickness is increased the 
contribution of the resonance states is enhanced. Nevertheless, those 
resonances located above the barrier (see the Fermi-Dirac wieghted curves in Fig.~\ref{fig7}) become less sensitive to barrier 
profile and hence the decline of TER ratio with barrier thickness increase.

\subsubsection{Metal/ CaO/BaTiO$_{3}$/Metal composite FTJ}

\begin{figure}[htp]
	\begin{center}
		\subfigure {\label{fig8a} \includegraphics [width=2.5in,height=2.5in]{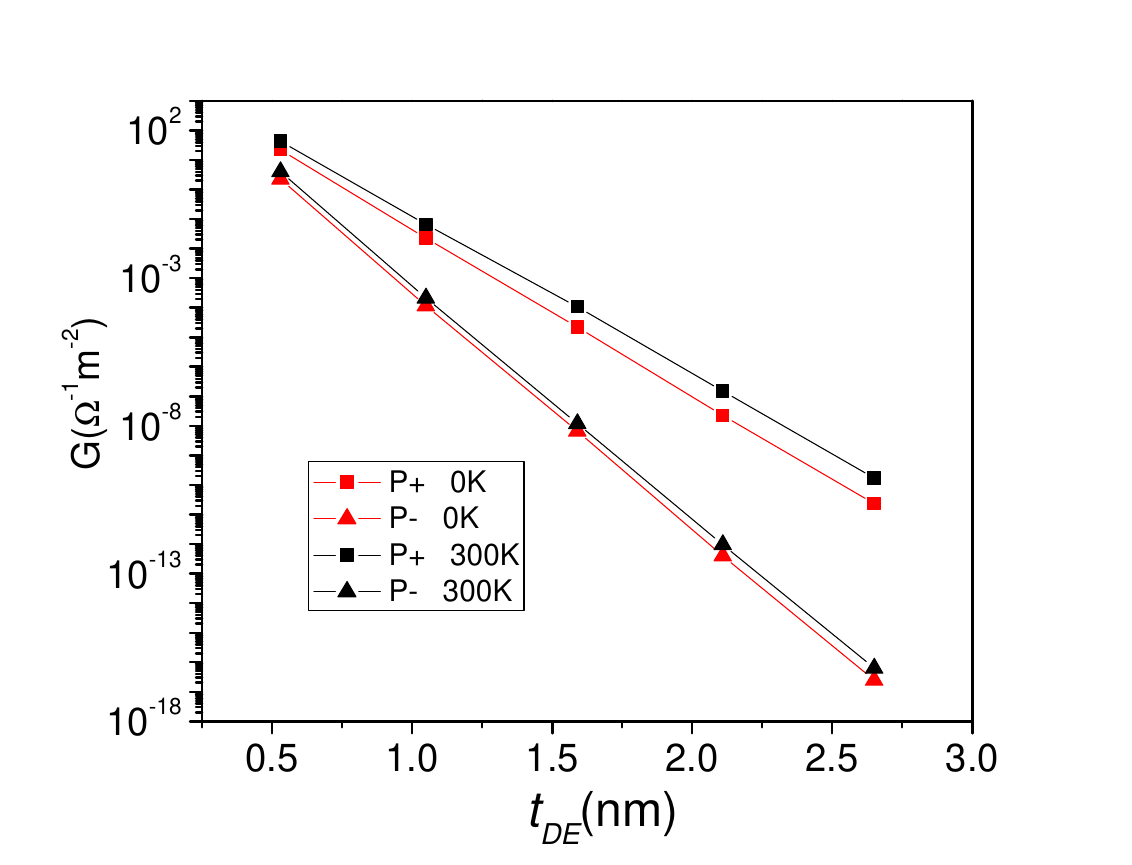}}  
		\subfigure {\label{fig8b} \includegraphics [width=2.50in,height=2.5in]{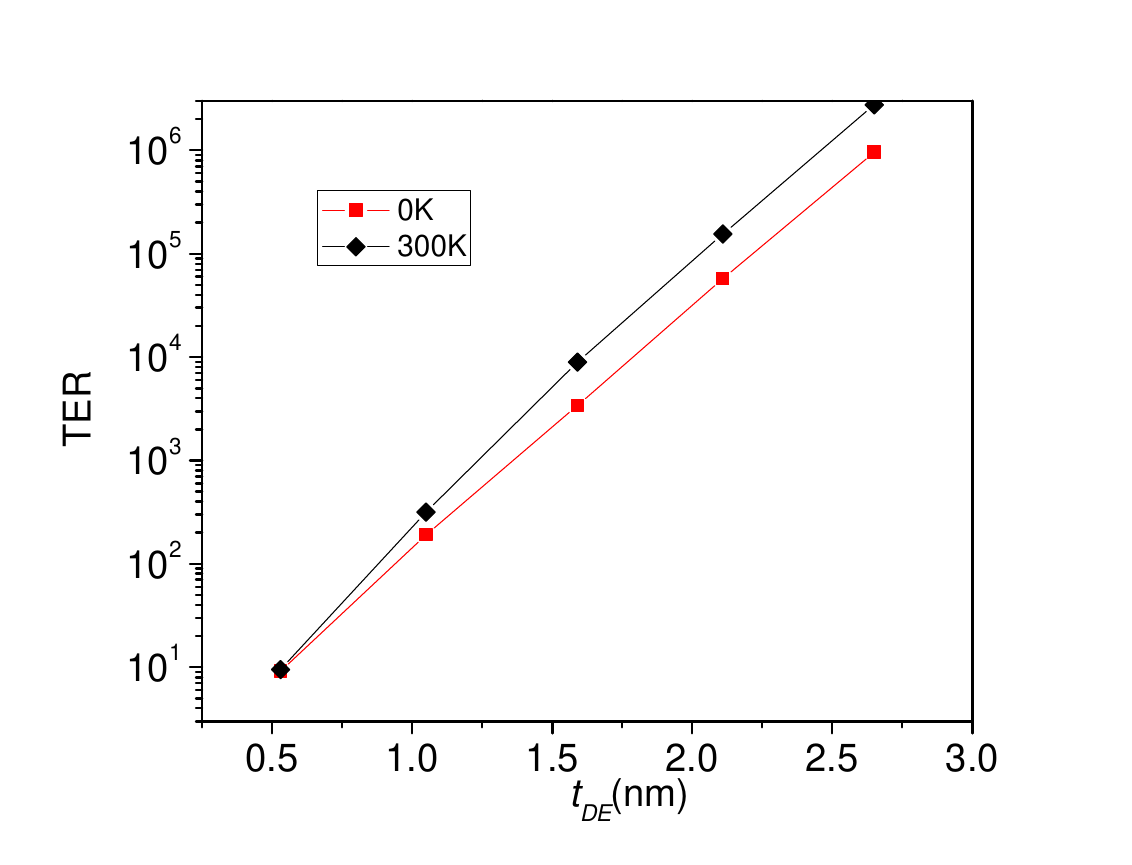}} 
	\end{center}
	\caption{Me/CaO/BaTiO$_{3}$/Me composite FTJ:(a) Conductance and (b) TER ratio as a function of the dielectric thickness at 0K and 300K.}
	\label{fig8}
\end{figure}

The third system studied herein is also a composite TFJ, with a CaO dielectric barrier
added to the BaTiO$_{3}$ ferroelectric barrier. The electrodes are of the 
same generic metal Me. In this case the asymmetry is ensured just by the presence of 
dielectric. We have set the the physical parameters to \cite{Zhuravlev2009}:
$\lambda_{1}=\lambda_{2}$=0.1 nm, $\varepsilon_{1}$= $\varepsilon_{2}$=1, $\varepsilon_{FE}$=90, $\varepsilon_{DE}$=10,, $E_{FL}=E_{FR}$=3 eV, $\phi_{1}$=5.5 
eV, $\phi_{2}$=3.6 eV, $\phi_{C}$=1.9 eV, $P$=40 $\mu $C/cm$^{2}$. The effective masses 
are equal to m$_{0}$ for all materials. We have also kept the thickness of 
BaTiO$_{3}$ to 2.4 nm and we have varied the thickness of CaO from 0.5 nm to 
3 nm. In comparison to the previous system, in this particular case the 
barrier is much higher on the dielectric side. The calculations of conductance 
and TER ratio are presented in Fig.\ref{fig8}.

\begin{figure}[htbp]
	\centerline{\includegraphics[width=2.5in,height=2.5in]{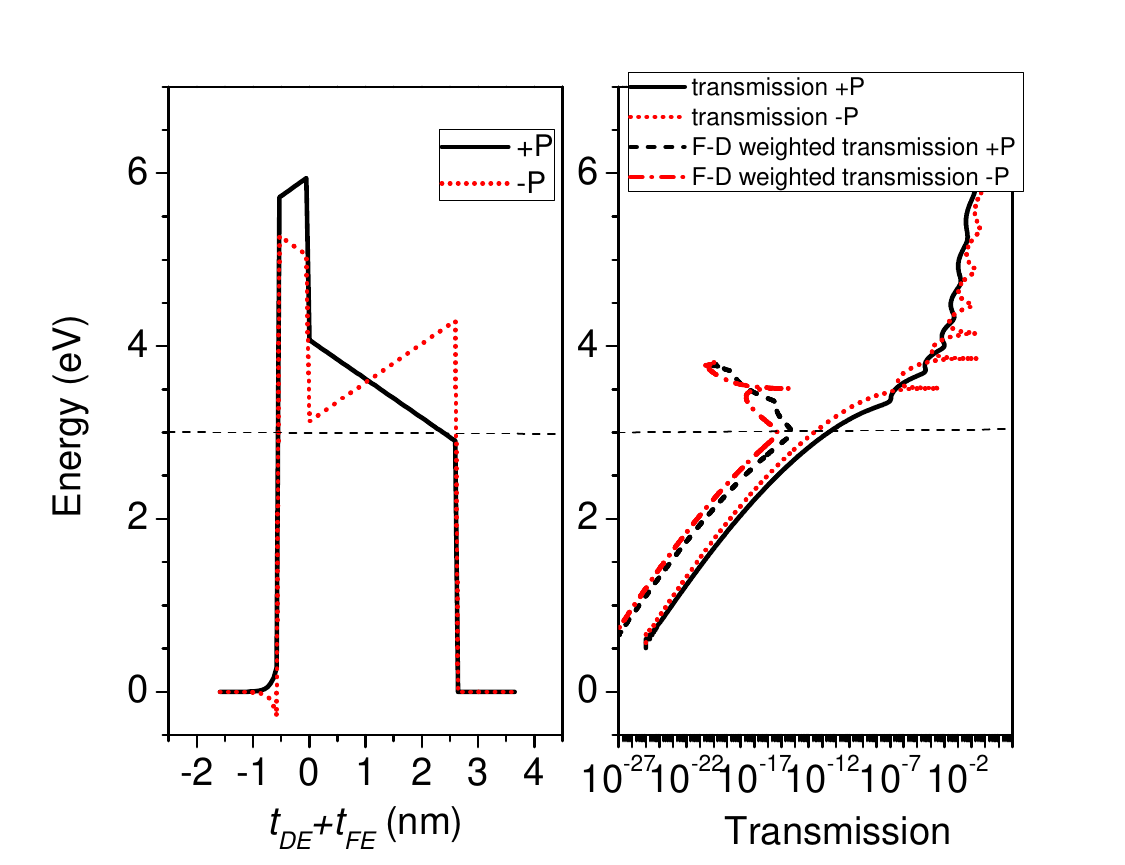}}
	\caption{Me/CaO/BaTiO$_{3}$/Me composite FTJ: the barrier profile, the 
		transmission probability coefficient, and the Fermi-Dirac weighted transmission probability at 300K; $t_{FE}$=2.5 nm, $t_{DE}$=0.5 nm. The dashed line in the graphs indicates the Fermi energy.}
	\label{fig9}
\end{figure}

In contrast to Pt/SrTiO$_{3}$/BaTiO$_{3}$/SrRuO$_{3}$, the conductance and the 
TER ratio of Me/CaO/BaTiO$_{3}$/Me composite FTJ exhibit an exponential 
dependence on dielectric thickness at both 0K and 300K. Therefore, 
qualitatively at least, a 0K analysis remains valid also at room 
temperature. In Fig.~\ref{fig9} we show the data for a quantitative explanation of conductance and 
TER ratio behavior with temperature. One can see that resonance states have 
a minor contribution to the current; the vast majority of carriers are 
transported through states around Fermi energy, although there are many 
resonances below the top of the barrier. Due to the higher dielectric 
barrier, these resonance states are just weakly coupled to the contacts, 
hence they show low transmission coefficients and they have a modest 
contribution to conductance. 

\subsection{The tunneling wavefunctions. The wavefunctions of resonances}

In the previous section it was discussed that the energy dependent transmission 
in FTJs has resonant features besides an exponential background. In this 
section we show that the wavefunctions also exhibit general features, some of 
which, e. g., those corresponding to resonances, are also observed in other 
nanostructures like quantum wells, multi-barrier structures, etc. Usually, 
as a scattering problem, the electron transport in FTJs has two solutions at 
a given energy: one solution for an incident electron wave coming from the 
left and the other for the electron wave coming from the right. In the 
following we will illustrate the wavefunctions at some representative energy 
values for two FTJs: Pt/BaTiO$_{3}$/SrRuO$_{3}$ and 
Pt/SrTiO$_{3}$/BaTiO$_{3}$/SrRuO$_{3}$

\subsubsection{The wavefunctions of Pt/BaTiO$_{3}$/SrRuO$_{3}$ FTJ}

\begin{figure}[htbp]
	\centerline{\includegraphics[width=2.5in,height=2.5in]{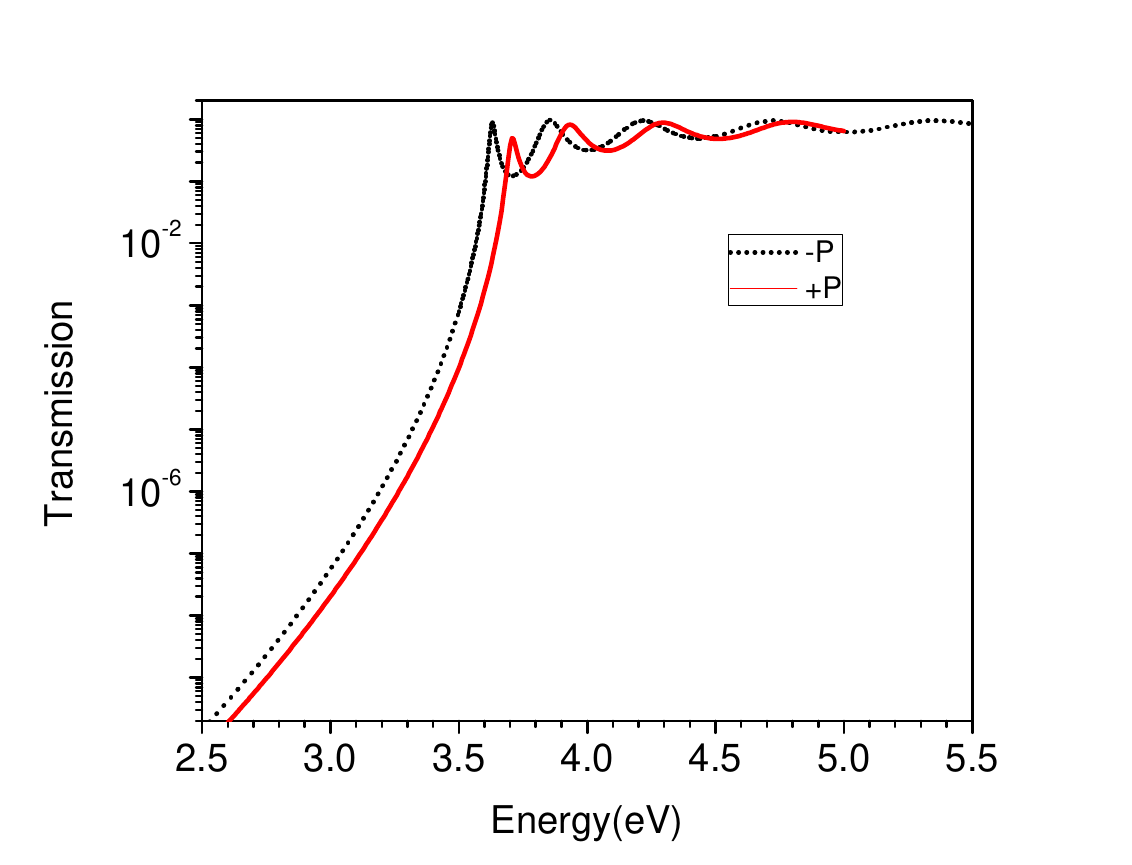}}
	\caption{Transmission probability coefficient for a BaTiO$_{3 }$ barrier of 1.6 nm 
		thickness between Pt and SrRuO$_{3}$ metallic contacts for both polarization directions.}
	\label{fig10}
\end{figure}

We illustrate the wave functions at Fermi energy (3 eV) and at some resonant energies like
that can be taken from Fig.~\ref{fig10}.

\begin{center}
\begin{figure}[htp]
	        \subfigure {\label{fig11a} \includegraphics   [width=2.5in,height=2.in]{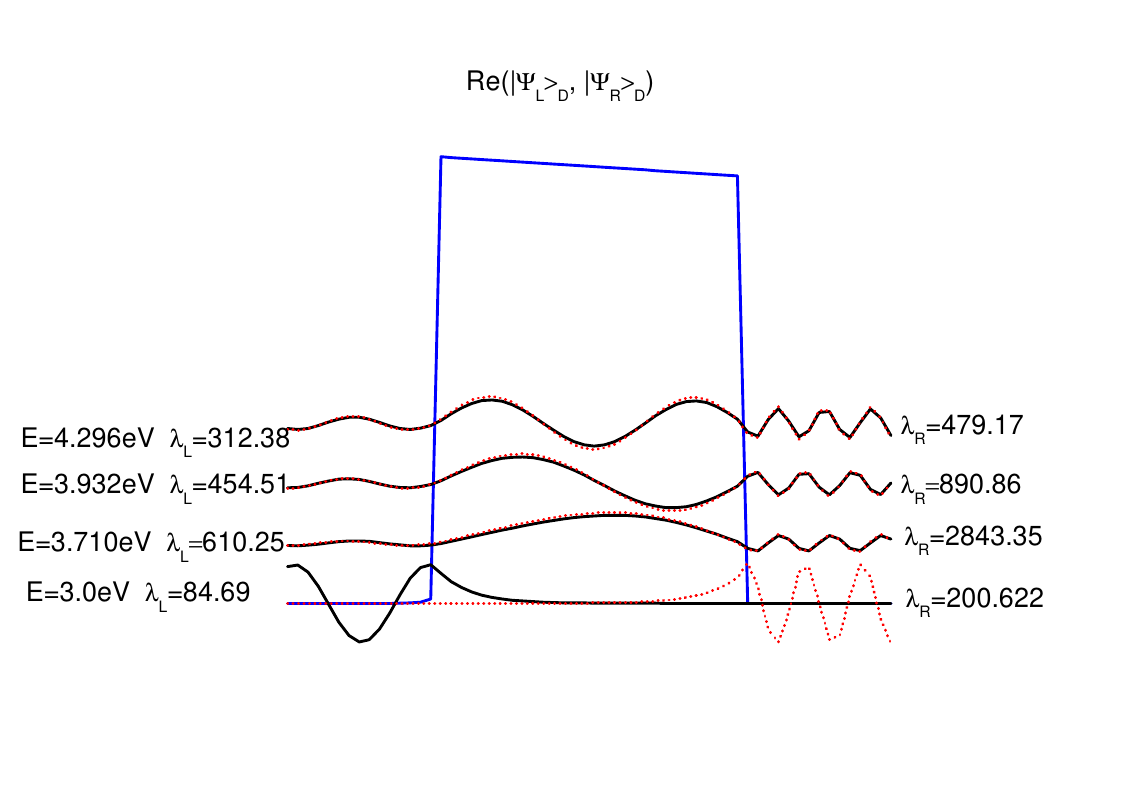}}  
	        \subfigure {\label{fig11b}  \includegraphics   [width=2.50in,height=2.in]{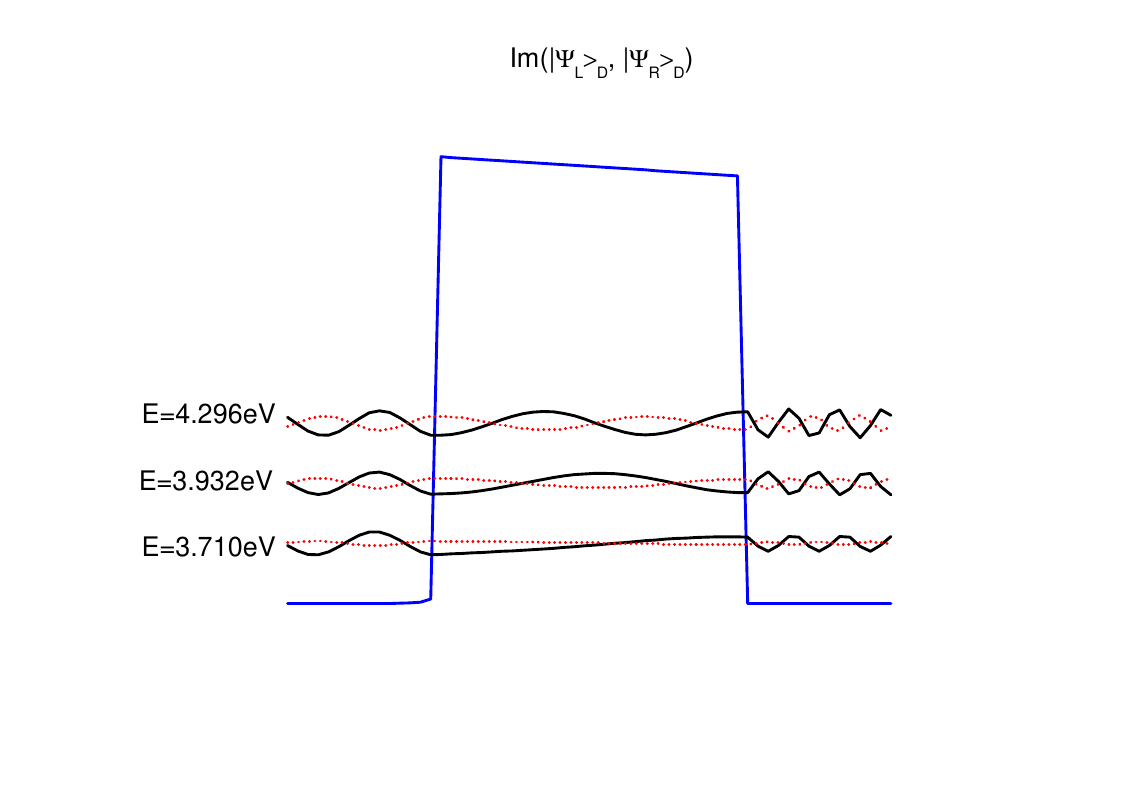}} 	\\
	        \subfigure {\label{fig11c}  \includegraphics   [width=2.5in,height=2.in]{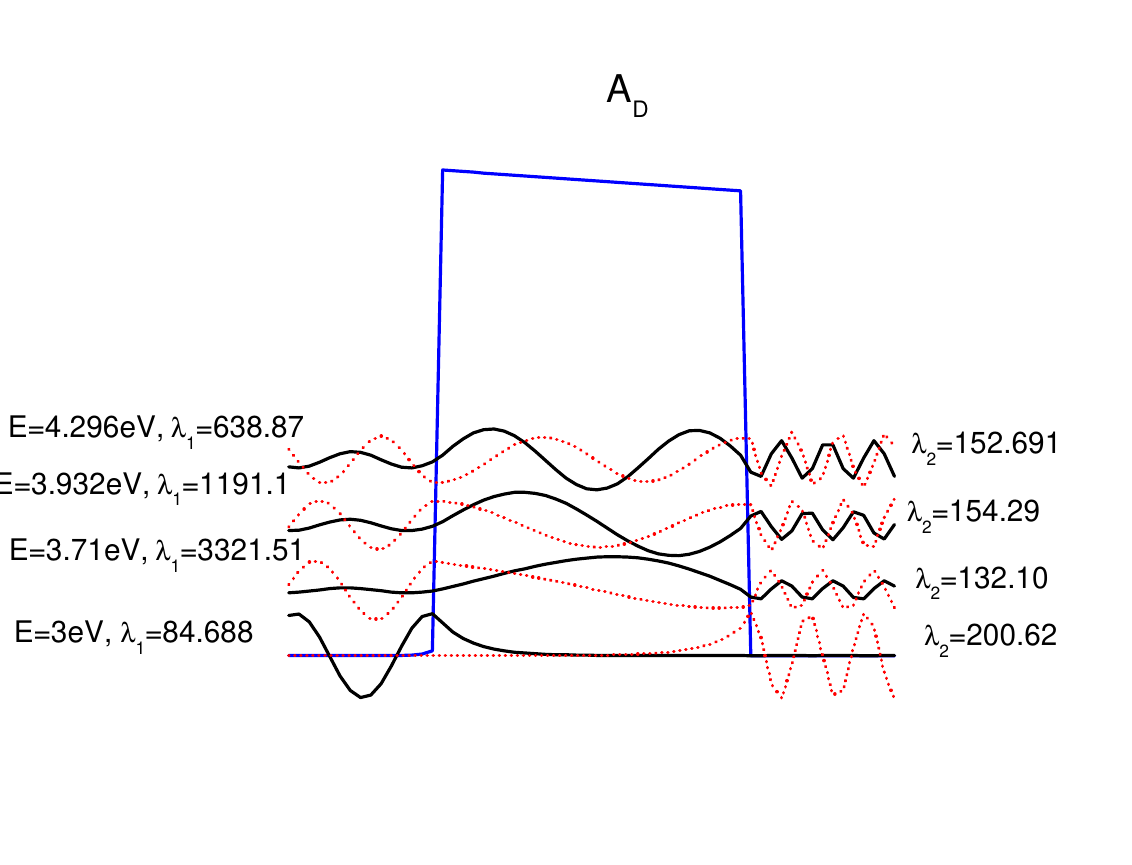}}  
		\subfigure {\label{fig11d} \includegraphics   [width=2.50in,height=2.in]{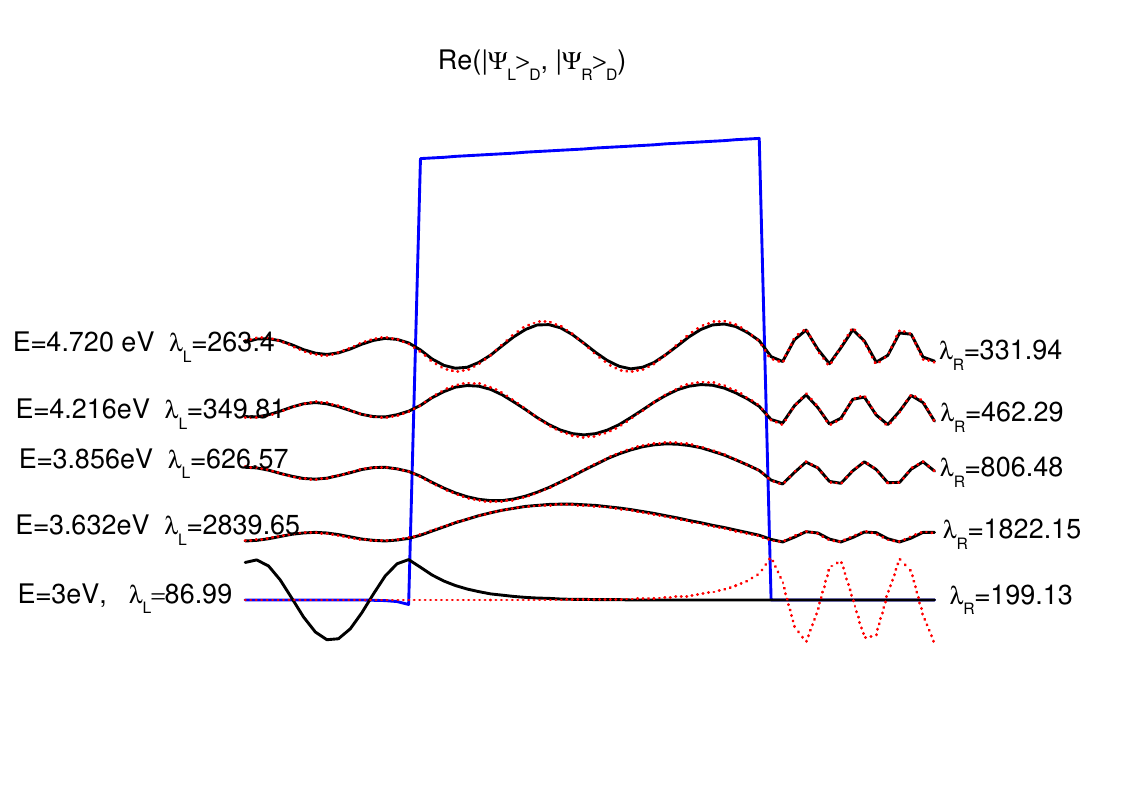}} \\ 
		\subfigure {\label{fig11e} \includegraphics   [width=2.5in,height=2.in]{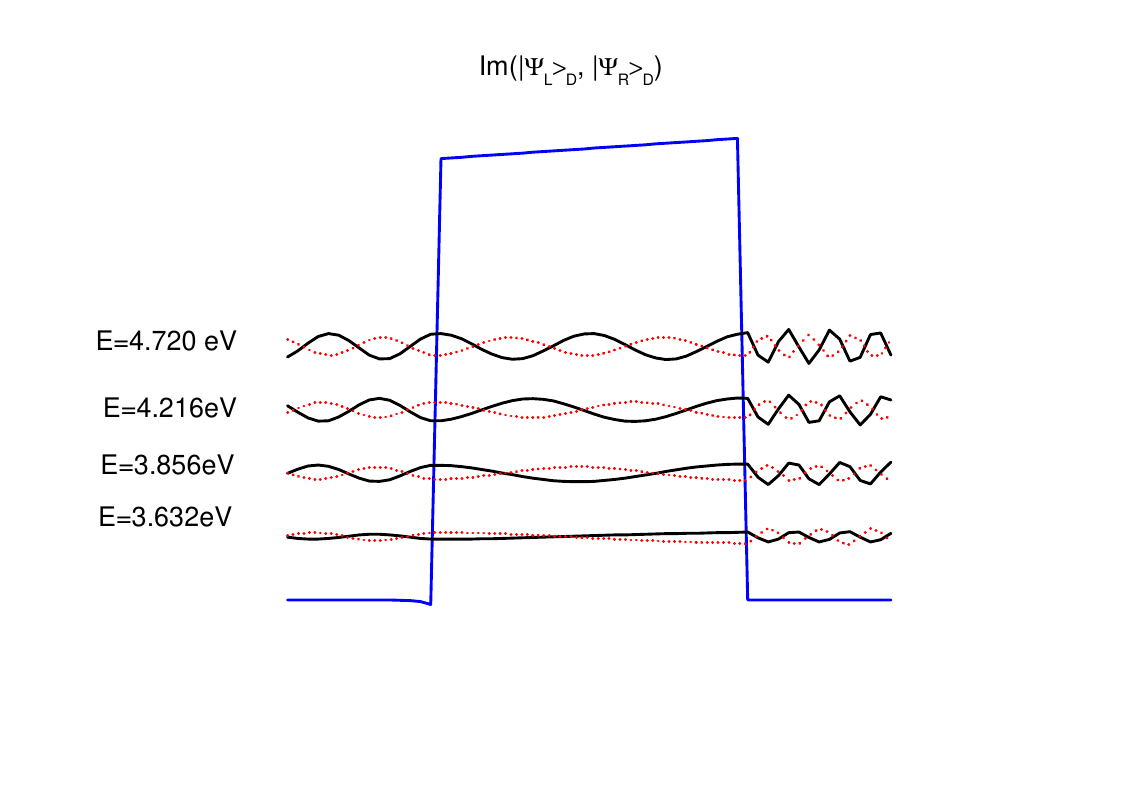}}  
		\subfigure {\label{fig11f} \includegraphics   [width=2.50in,height=2.in]{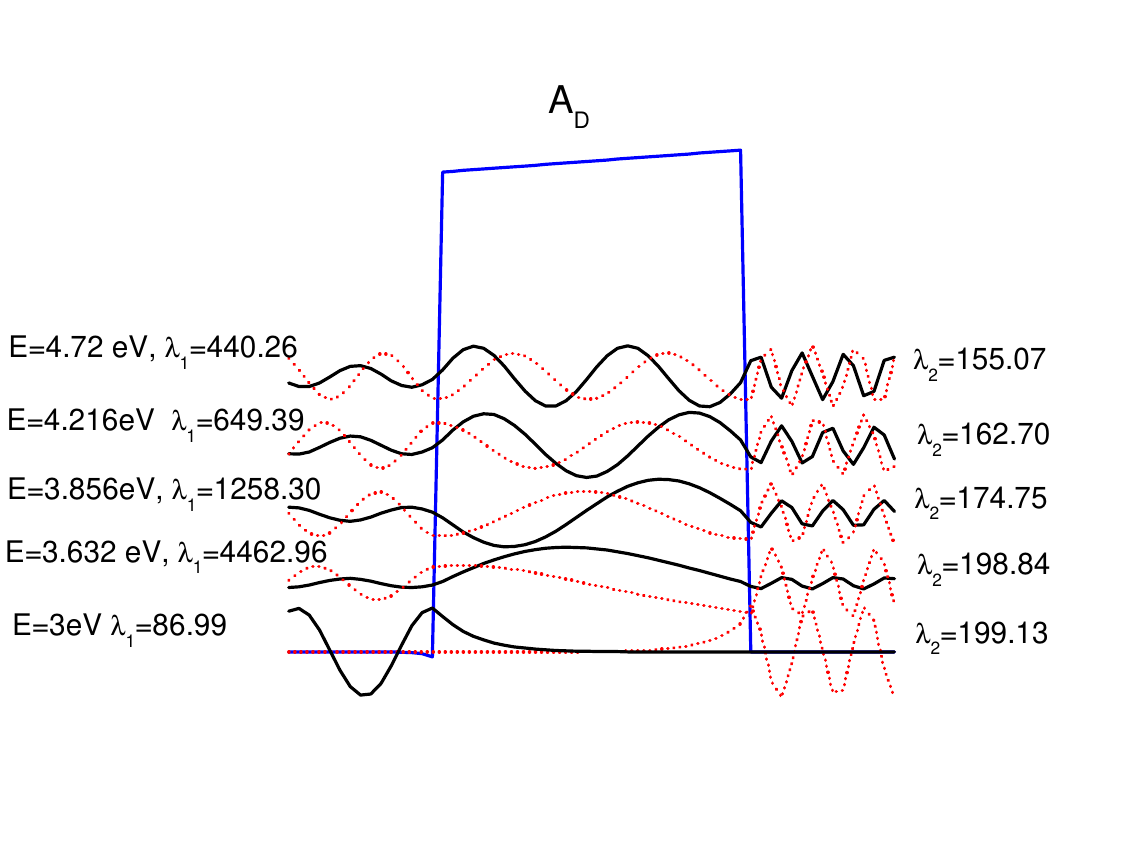}} 
	
	\caption{Representative wavefunctions of BaTiO$_{3 }$FTJ as normalized 
		eigenvectors of $A_{L}$ (solid black line), $A_{R}$ (dotted red line). They 
		are scaled down by a factor $\sqrt {{\lambda _{L,R} } \mathord{\left/ 
				{\vphantom {{\lambda _{L,R} } {2\pi }}} \right. \kern-\nulldelimiterspace} 
			{2\pi }} $, where \textit{$\lambda $}$_{L,R}$ is the corresponding eigenvalues of $A_{L}$, 
		$A_{R}$. In addition we illustrate the normalized eigenvectors of $A_{D}$ with 
		eigenvalue \textit{$\lambda $}$_{1}$ (solid black line) and \textit{$\lambda $}$_{2}$ (dotted red line). The 
		normalized eigenvectors of $A_{L}$, $A_{R}$ are as follows: (a) - real part, positive 
		ferroelectric polarization; (b) - imaginary part, positive ferroelectric 
		polarization; (d) - real part, negative ferroelectric polarization; (e) - 
		imaginary part, negative, ferroelectric polarization. The normalized 
		eigenvectors of $A_{D}$, are as follows: (c) - positive ferroelectric polarization; 
		(f) - negative ferroelectric polarization. }
		
	\label{fig11}
\end{figure}
\end{center}

\noindent
It is obvious that up to about 3.5 eV the transmission probability has an 
exponential dependence with respect to energy. In Fig.~\ref{fig11} we plotted the 
wavefunctions of a 1.6 nm thick BaTiO$_{3}$ barrier at Fermi energy as well 
as the first few resonance energies for both directions of polarization. In 
order to compare those wavefunctions we plotted the ``normalized'' 
eigenvectors of $A_{L}$, $A_{R}$, and $A_{D}$. In other words, for instance, 
$\left| {\Psi _{L,R} } \right\rangle _D $ are divided by $\sqrt {{\lambda 
		_{L,R} } \mathord{\left/ {\vphantom {{\lambda _{L,R} } {2\pi }}} \right. 
		\kern-\nulldelimiterspace} {2\pi }} $ and the eigenvectors of $A_{D}$ are 
divided by $\sqrt {{\lambda _{1,2} } \mathord{\left/ {\vphantom {{\lambda 
					_{1,2} } {2\pi }}} \right. \kern-\nulldelimiterspace} {2\pi }} $ with 
$\lambda _{1,2} $, the corresponding eigenvalues. The device region is comprised of 
the barrier and several layers of contact regions, such that the region 
outside the device should exhibit a flat electrostatic potential. In 
practice this is achieved when a few nanometers of contacts are added to 
device region. At Fermi energy the wavefunctions are almost real and their 
overlap is almost zero. They decay exponentially in the barrier, hence the 
Simmons or Brinkman formula applies for states around Fermi energy. 

Since the barrier is thin the resonances are well separated. From Figs.~\ref{fig11a},~\ref{fig11b},
and Figs.~\ref{fig11d},~\ref{fig11e} we see that the ``normalized'' $\left| {\Psi _L } \right\rangle _D $ is almost identical 
with the complex conjugation of ``normalized''$\left| {\Psi _R } 
\right\rangle _D $, however the values of ${\lambda _{L,R} } \mathord{\left/ 
	{\vphantom {{\lambda _{L,R} } {2\pi }}} \right. \kern-\nulldelimiterspace} 
{2\pi }$ provide the levels of electron density inside the device for the 
corresponding wavefunction. These states exhibit a confining character 
within the barrier, in contrast to the states shown at Fermi energy. These 
solutions belong to resonance and anti-resonance states in the complex 
wavevector plane \cite{Calderon2010,Tolstikhin1997,Hatano2011}. Moreover, the real and the imaginary parts of 
$\left| {\Psi _{L,R} } \right\rangle _D $ can be found in the normalized 
eigenvectors of $A_{D}$ (Figs.~\ref{fig11c} and \ref{fig11f}). Finally, by analyzing Figs.~\ref{fig5} 
and \ref{fig11}, we notice that just the first resonance would participate to the 
electron transport at room temperature.

\subsubsection{The wavefunctions of Pt/SrTiO$_{3}$/BaTiO$_{3}$/SrTiO$_{3}$ composite FTJ}

\begin{figure}[htp]
	\begin{center}
	    \subfigure {\label{fig12a} \includegraphics   [width=2.5in,height=2.in]{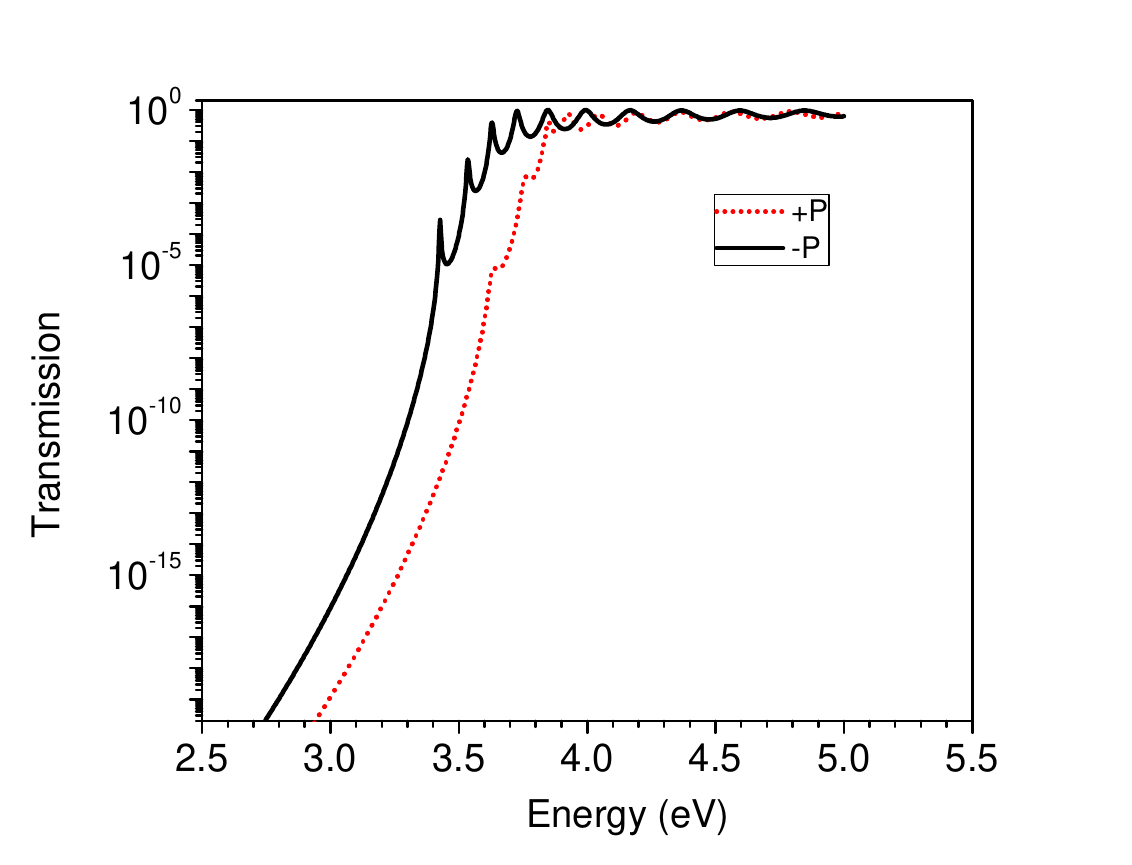}}  
            \subfigure {\label{fig12b} \includegraphics   [width=2.50in,height=2.in]{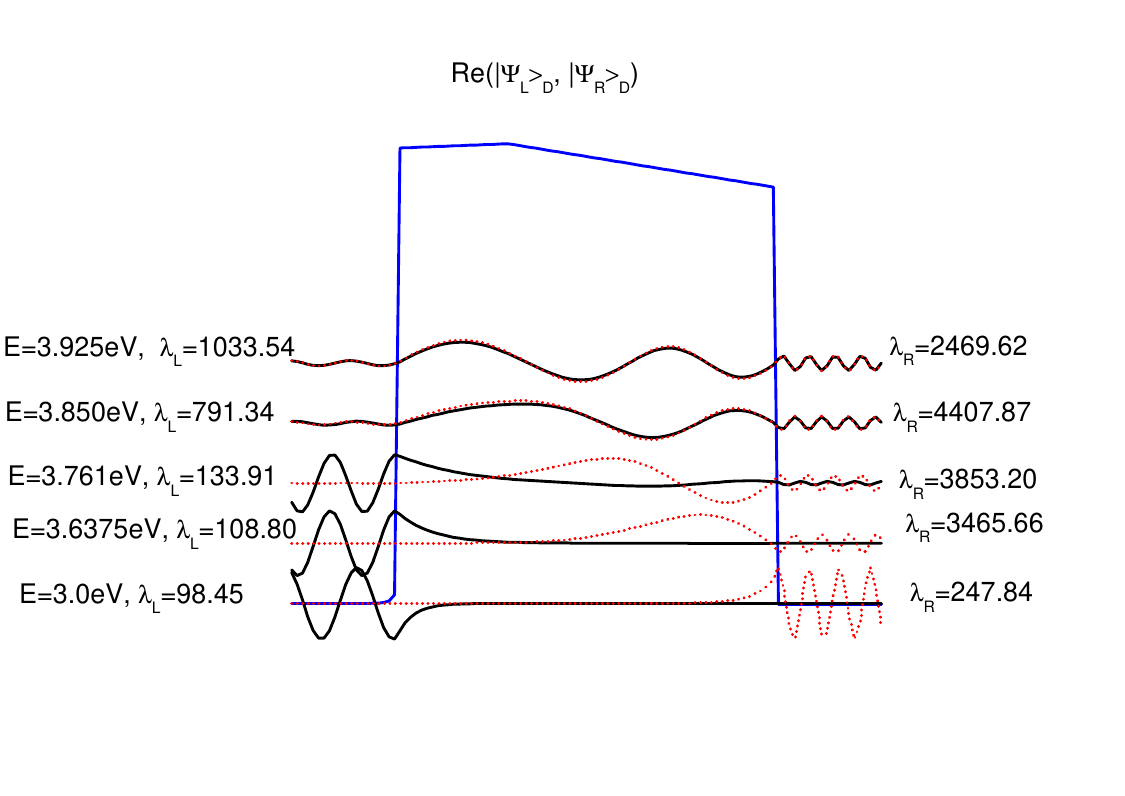}} 	
	    \subfigure {\label{fig12c} \includegraphics   [width=2.5in,height=2.in]{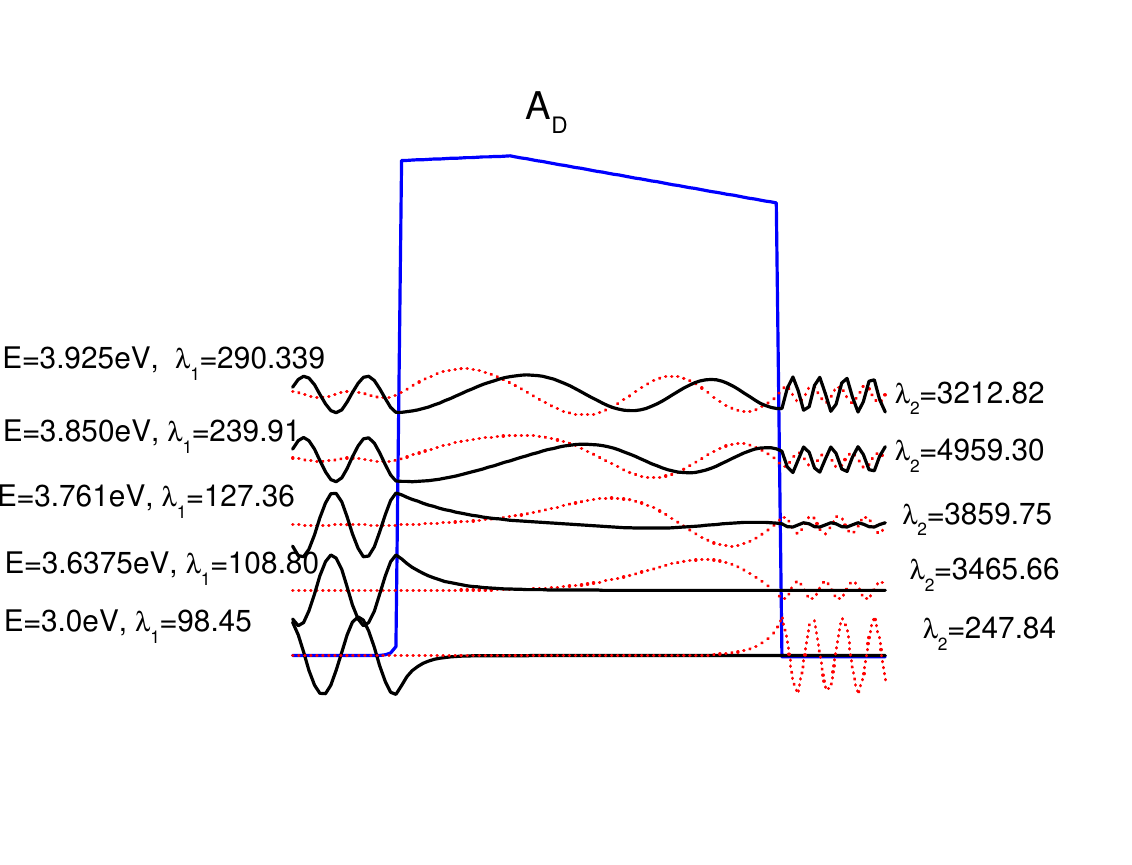}}  
            \subfigure {\label{fig12d} \includegraphics   [width=2.50in,height=2.in]{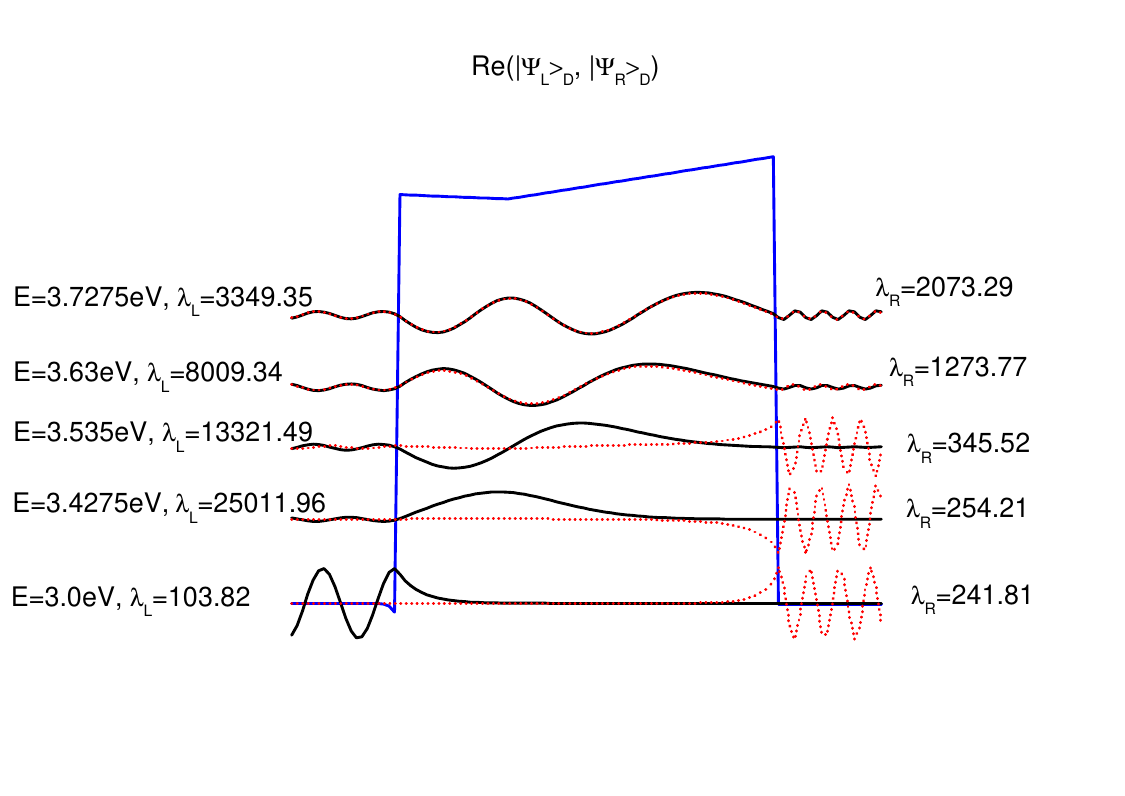}}  
	    \subfigure {\label{fig12e} \includegraphics   [width=2.5in,height=2.in]{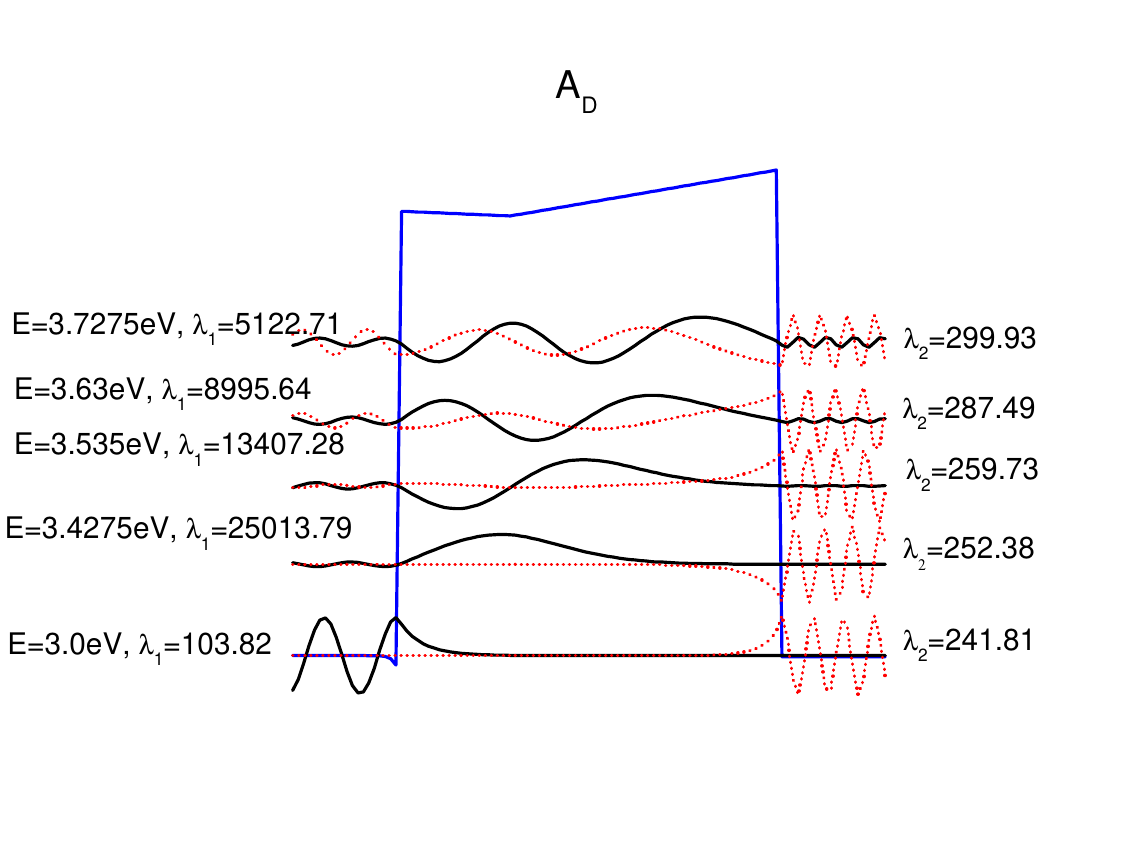}}
          
	\end{center}
	\caption{Pt/SrTiO$_{3}$/BaTiO3/ SrTiO$_{3}$ composite FTJ (1 nm of 
		SrTiO$_{3}$ and 2.4 nm of BaTiO$_{3})$. (a) Transmission probability for 
		positive (dotted red line) and negative (solid black line) ferroelectric 
		polarization; (b)-(e) Like in Fig.~\ref{fig11} the representative wavefunctions are 
		the normalized eigenvectors of $A_{L}$ and $A_{R}$ as well as the normalized 
		eigenvectors of $A_{D}$. The real part of normalized eigenvectors of 
		$A_{L}$ (solid black line) $A_{R}$ (dotted red line) are as follows: (b) -- positive 
		ferroelectric polarization; (d) - negative ferroelectric polarization. The 
		normalized eigenvectors of $A_{D}$ with eigenvalue \textit{$\lambda $}$_{1}$ (solid black line) 
		and \textit{$\lambda $}$_{2}$ (dotted red line) are as follows: (c) - positive ferroelectric 
		polarization; (e) - negative ferroelectric polarization.  }
	
	\label{fig12}
\end{figure}

The plots of the wave functions for this composite FTJ are shown in Fig.~\ref{fig12}. 
The transmission probability for polarizations is shown in Fig.~\ref{fig12a} from 
which we can extract the resonances that play a role in the electron 
transport at room temperature. The physical parameters are those that have 
already been used in the calculations. The effective thickness of the 
barrier is larger than in simple FTJ presented in the previous subsection; 
hence the resonances are closely spaced. Still, at Fermi energy the 
wavefunctions decay exponentially in the barrier and are almost real, while 
their overlap is almost zero. The wavefunctions of the first four resonances 
in transmission have a much smaller imaginary part, which is not shown here. 
Like in the previous section, these wavefunctions manifest confining 
character in the barrier. Moreover, the dielectric induces a much smaller 
coupling to one of the two contacts, such that the first two resonances in 
transmission contain a significant background contribution as a decaying 
wavefunction in the barrier seen in the solution from the left (Fig.~\ref{fig12b}, 
positive polarization) and in the solution from the right (Fig.~\ref{fig12d}, 
negative polarization). Analyzing Figs~\ref{fig12c} and \ref{fig12e} one can notice that the 
overlap between the solution from the left and that from the right is almost 
zero for the first two resonances in transmission. The next two resonances 
in transmission, however, can be distinguished from the background, the 
normalized eigenvectors of $A_{L}$ and $A_{R}$, being quite similar, yet the 
corresponding wavefunctions $\left| {\Psi _L } \right\rangle _D $ and 
$\left| {\Psi _R } \right\rangle _D $ have quite different amplitudes. The 
difference in amplitudes starts to close in as we move to higher energies, 
since at sufficiently higher energy the asymmetry of the barrier will not play 
any major role in the electron transmission. As a final comment, all four 
resonances shown here participate to temperature activated transport (Fig.~\ref{fig7}).
However, the main contribution to temperature dependent transport is 
given by the first three resonances, even though the couplings to the 
contacts of the first two resonances are not so strong as the couplings of 
the third one which is further apart in energy.

\section{Conclusions}

In conclusion, the semi-empirical model of electrostatic and NEGF 
calculations can provide a detailed picture of electron transport in systems 
with FTJs. It treats on equal footing several transport mechanisms that are 
usually invoked when studying these devices like direct tunneling, 
thermionic emission, and Fowler-Nordheim tunneling. This feature of NEGF 
allows us to assess in detailed form the role of temperature in electron 
transport and how temperature affect TER ratio. We have found that for 
simple or composite BaTiO$_{3}$ based FTJs the transport through temperature 
activated resonance (Gamow-Siegert) states may become dominant, affecting 
both the conductance and the TER ratio; thus, the more resonances are 
activated and the stronger their couplings to the contacts the more powerful 
is the effect of temperature. The effect of temperature is obvious in 
thicker FTJs, since the resonance states, especially those above the 
barrier, are closer to each other, hence more of them may 
participate in the transport. More insides into this phenomenology may be acquired 
by calculating 
and plotting the wavefunctions at any given energy. We show that this is 
possible for NEGF calculations of single-band transport. Thus the transport by direct 
tunneling is made through states whose wavefunctions have a decaying shape 
in the barrier. These states belong to the background generated by all 
resonance states. Furthermore, where the resonances are strong and well 
separated, the wavefunctions show confinement character in the barrier. In 
the intermediate regime, where the resonances are weak, the confining 
character of the resonance competes with the decaying character of the 
background. Lastly, we suggest that these thorough insights can be used in the 
optimization process of the FTJ design in various applications, particularly at high 
temperature.

\appendix
\section{The BenDaniel and Duke Hamiltonian}
The BenDaniel and Duke Hamiltonian 

\begin{equation}
	\label{A1}
H = - \frac{\hbar ^2}{2}\frac{d}{dx}\left( {\frac{1}{m^\ast \left( x 
		\right)}\frac{d}{dx}} \right) + \frac{\hbar ^2k^2}{2m^\ast \left( x \right)} 
+ U\left( x \right)
\end{equation}
\noindent
can be cast into 1D problem by rewriting it as 

\begin{equation}
	\label{A2}
H = - \frac{  \hbar ^{2}}{{2}}\frac{d}{dx}\left( 
{\frac{{1}}{m^\ast \left( x \right)}\frac{d}{dx}} \right) + V_k (x) + 
\frac{\hbar ^{2}k^2}{{2}m_L^\ast },
\end{equation}

\noindent
where $m_L^\ast $ is the effective mass in the left contact and 

\begin{equation}
	\label{A3}
V_k \left( x \right) = U\left( x \right) + \frac{\hbar 
	^{2}k^2}{{2}m_L^\ast }\left( {\frac{m_L^\ast }{m^\ast \left( x 
		\right)} - {1}} \right).
\end{equation}

Equation (\ref{A2}) is discretized and its discrete form can be mapped 
into a tight-binding Hamiltonian. Thus the continuous variable $x$ is 
transformed in the discrete version \textit{$\Delta \cdot $n}, where $\Delta $ is the discretization 
step and $n$ is an integer running from -$\infty $ to $\infty $. We define a 
localized orbital $\left| {n,R^L} \right\rangle $ with \textit{$\Delta \cdot $n} and $R^{L}$ as 
longitudinal and transverse positions. Moreover, we construct transverse 
Bloch orbitals as a sum over $N$ localized orbitals in the transverse plane 

\begin{equation}
	\label{A4}
\left| {n,k} \right\rangle = \frac{\mbox{1}}{\sqrt N }\sum\limits_{R^L} 
{e^{ik\mbox{R}^L}} \left| {n,R^L} \right\rangle .
\end{equation}

In this Bloch basis the Hamiltonian has the following expression 

\begin{equation}
	\label{A5}
\left\langle {n,k} \right|H\left| {n\mbox{',}k} \right\rangle = D_n \left( k 
\right)\delta _{n,n'} - t_{n,n'} \delta _{n\mbox{',}n\pm 1} ,
\end{equation}
\noindent
where 

\begin{equation}
	\label{A6}
D_n \left( k \right) = \frac{\hbar ^{2}}{{2}\Delta 
	^{2}}\left( {\frac{1}{m^ - } + \frac{1}{m^ + }} \right) + 
V_k \left( n \right),
\end{equation}

\begin{equation}
	\label{A7}
t_{n,n'} = \frac{\hbar ^{2}}{\left( {m_n^\ast + m_{n'}^\ast } 
	\right)\Delta ^{2}},
\end{equation}

\begin{equation}
	\label{A8}
m^ - = \frac{m_{n - {1}}^\ast + m_n^\ast }{2},
\end{equation}

\begin{equation}
	\label{A9}
m^ + = \frac{m_n^\ast + m_{n + 1}^\ast }{2}.
\end{equation}

The tight-binding version (\ref{A5}) of BenDaniel and Duke Hamiltonian has a 
tridiagonal form. The left contact is defined for $n$ running from -$\infty $ 
to 0, the device is defined for $n $ running from 1 to $n_{D}$, and the right 
contact from $n_{D}$ +1 to $\infty $. 
The left and the right contacts are homogeneous systems; hence we define the 
tight-binding parameters as follows. The diagonal term of the left (right) 
contact is defined as $D_{L}$ ($D_{R})$, while the off-diagonal terms 
$t_{i,i\pm 1}$ as $t_{L}$ ($t_{R})$.

\begin{acknowledgments}
  The authors acknowledge financial support from UEFISCDI Grant No. PN-III-P4-ID-PCE-2020-1985 and from the Romanian Core Program Contract No.14 N/2019 Ministry of Research, Innovation, and Digitalization.
\end{acknowledgments}

\bibliography{nanomat}

\end{document}